\documentclass[12pt]{spieman}  % 12pt font required by SPIE;
\usepackage{amsmath,amsfonts,amssymb}
\usepackage{graphicx}
\usepackage{setspace}
\usepackage{tocloft}
\usepackage{multirow}
\usepackage{caption}
\usepackage{lineno}
\usepackage{xcolor}

\title{Technical description and performance of the phase II version of the Keck Planet Imager and Characterizer}

\author[a,*]{Nemanja~Jovanovic}
\author[a]{Daniel~Echeverri}
\author[b]{Jacques-Robert~Delorme}%[0000-0001-8953-1008]
\author[c]{Luke~Finnerty} %[0000-0002-1392-0768]
\author[a]{Tobias~Schofield}
\author[d]{Jason~J.~Wang}
\author[a]{Yinzi~Xin}
\author[a]{Jerry~Xuan}%[0000-0002-6618-1137]
\author[e]{J.~Kent~Wallace}
\author[a,e]{Dimitri~Mawet}
\author[a,h]{Aniket Sanghi}

\author[a]{Ashley~Baker}

\author[e]{Randall~Bartos}

\author[f]{Charlotte~Z.~Bond}

\author[c]{Benjamin~Calvin}

\author[b]{Sylvain~Cetre}

\author[b]{Greg~Doppmann}

\author[c]{Michael~P.~Fitzgerald}
\author[a]{Jason~Fucik}

\author[g]{Maodong Gao}

\author[g]{Jinhao Ge}

\author[b]{Charlotte Guthery}

\author[a,h]{Katelyn Horstman}
\author[d]{Chih-Chun Hsu}%[0000-0002-5370-7494]

\author[a,l]{Joshua~Liberman}
\author[g]{Stephanie Leifer}

\author[b]{Scott Lilley}

\author[c]{Ronald~Lopez}

\author[b]{Eduardo Marin}

\author[i]{Emily~C.~Martin}
\author[e]{Bertrand~Mennesson}

\author[i]{Evan~Morris}

\author[a]{Reston~Nash}

\author[a]{Jacklyn~Pezzato}

\author[a]{Michael~Porter}

\author[a]{Mitsuko~Roberts}

\author[e]{Garreth~Ruane}
\author[j]{Jean-Baptiste~Ruffio}
\author[j]{Ben~Sappey}

\author[e]{Eugene~Serabyn}

\author[g]{Boqiang~Shen}

\author[i]{Andrew~Skemer}

\author[k]{Ji~Wang}
\author[b]{Edward Wetherell}

\author[b]{Peter~Wizinowich}

\author[i]{Ma\"issa Salama}%[0000-0002-5082-6332]

\author[i]{Vincent Chambouleyron}

\author[i]{Rebecca Jensen-Clem}%[0000-0003-0054-2953

\author[m]{Chas Beichman}%[0000-0003-0054-2953

\affil[a]{Department of Astronomy, California Institute of Technology, Pasadena, CA 91125, USA}
\affil[b]{W. M. Keck Observatory, 65-1120 Mamalahoa Hwy, Kamuela, HI, USA}
\affil[c]{Department of Physics \& Astronomy, 430 Portola Plaza, University of California, Los Angeles, CA 90095, USA}
\affil[d]{Center for Interdisciplinary Exploration and Research in Astrophysics (CIERA) and Department of Physics and Astronomy, Northwestern University, Evanston, IL 60208, USA}
\affil[e]{Jet Propulsion Laboratory, California Institute of Technology, 4800 Oak Grove Dr., Pasadena, CA 91109, USA}
\affil[f]{UK Astronomy Technology Centre, Royal Observatory, Edinburgh EH9 3HJ, United Kingdom}
\affil[g]{Department of Applied Physics, California Institute of Technology, Pasadena, CA, United States of America}
\affil[h]{NSF Graduate Research Fellow}
\affil[i]{Department of Astronomy \& Astrophysics, University of California, Santa Cruz, 1156 High Street, Santa Cruz, CA 95064, USA}
\affil[j]{Department of Astronomy \& Astrophysics,  University of California, San Diego, La Jolla, CA 92093, USA}
\affil[k]{Department of Astronomy, The Ohio State University, 100 W 18th Ave, Columbus, OH 43210 USA}
\affil[l]{James C. Wyant College of Optical Sciences, University of Arizona, Meinel Building 1630 E. University Blvd., Tucson, AZ. 85721}
\affil[m]{NASA Exoplanet Science Institute/IPAC, Jet Propulsion Laboratory, California Institute of Technology, 1200 E California Boulevard, Pasadena, CA 91125, USA}

\cftpagenumbersoff{figure}
\cftpagenumbersoff{table} 
\begin{document} 
%\linenumbers
\maketitle

\begin{abstract}
The Keck Planet Imager and Characterizer (KPIC) is a series of upgrades for the Keck II Adaptive Optics (AO) system and the NIRSPEC spectrograph to enable diffraction-limited, high-resolution ($R>30,000$) spectroscopy of exoplanets and low-mass companions in the K and L bands. Phase I consisted of single-mode fiber injection/extraction units (FIU/FEU) used in conjunction with an H-band pyramid wavefront sensor. The use of single-mode fibers provides a gain in stellar rejection, a substantial reduction in sky background, and an extremely stable line-spread function in the spectrograph. Phase II, deployed and commissioned in 2022, brought a 1000-actuator deformable mirror, beam-shaping optics, a vortex mask, and other upgrades to the FIU/FEU. An additional service mission in 2024 extended operations down to y band, delivered an atmospheric dispersion corrector, and provided access to two laser frequency combs. KPIC phase II brings higher planet throughput, lower stellar leakage and many new observing modes which extend its ability to characterize exoplanets at high spectral resolution, building on the successes of phase I.

In this paper we present a description of the final phase II version of KPIC, along with results of system-level laboratory testing and characterization showing the instrument’s phase II throughput, stability, repeatability, and other key performance metrics prior to delivery and during installation at Keck. We outline the capabilities of the various observing modes enabled by the new modules as well as efforts to compensate for static aberrations and non-common path errors at Keck, which were issues that plagued phase I. Finally, we show results from commissioning.

\end{abstract}

% Include a list of up to six keywords after the abstract

% Include email contact information for corresponding author
{\noindent \footnotesize\textbf{*}Nemanja Jovanovic,  \linkable{nem@caltech.edu} }

\begin{spacing}{1}   % use double spacing for rest of manuscript

%%%%%%%%%%%%%%%%%%%%%%%%%%%%%%%%%
\section{Introduction}
The Keck Planet Imager and Characterizer (KPIC) is an instrument designed to enable high spectral resolution characterisation of directly imaged exoplanets and brown dwarfs in the near-infrared~\cite{Mawet2017_KPIC}. It combines high contrast imaging (HCI) with high resolution spectroscopy (HRS)~\cite{Wang2017_HDC,Mawet2017_HDCII}. This combination of techniques was first proposed by~\citenum{Sparks-ISE2002} and later refined by~\citenum{Riaud-IEL2007} and~\citenum{Snellen2015_HDC}. This approach relies on a high contrast imaging system to first spatially isolate the planet light and suppress the starlight. Then, a high resolution spectrometer is used to differentiate the signals between the two components based on differences in properties such as composition and radial velocity. KPIC, which is deployed at the Keck Observatory, was the first dedicated instrument optimized for this observational technique. The facility adaptive optics (AO) system first corrects for the atmosphere, and then light from the directly-imageable companion is injected into a single-mode fiber (SMF) and routed to NIRSPEC~\cite{McLean1998_NIRSPEC,McLean2000_NIRSPEC,Martin2018}, a high resolution spectrometer ($R=\lambda/\Delta\lambda>30,000$, where $\lambda$ is the wavelength and $\Delta\lambda$ is the resolution). In this way, KPIC is an interface between Keck AO and NIRSPEC to enable HCI+HRS. The SMFs not only provide a convenient way to transport the light to the spectrometer, but more importantly, they act as a spatial filter, helping to minimize stellar leakage as proposed by~\citenum{Mawet2017_HDCII}.   

KPIC was deployed in phases, with each phase bringing additional sub-systems and modules to the instrument. Phase I was deployed in 2018, and brought the core elements needed for fiber injection, control, and fiber extraction. This included a front end with a tracking camera working in concert with a tip/tilt mirror to steer the beam and inject it into a bundle of SMFs, as well as a back-end with re-imaging optics to project the coupled fiber light onto the slit of NIRSPEC. %The front-end dealing with with coupling light into the fibers is referred to as the Fiber Injection Unit (FIU), while the back-end which interfaces the bundle to NIRSPEC is the Fiber Extraction Unit (FEU). 
Phase I also included a near-infrared (NIR) pyramid wavefront sensor (PyWFS) to improve wavefront correction and enable guiding on fainter targets~\cite{Bond2020_PyWFS}. The design, testing, and deployment of KPIC phase I was described in great detail in earlier works~\cite{Delorme2021_KPIC, Morris2020_KPICPhaseI}. That minimalist version of KPIC was commissioned between 2018 and 2020 and proved to be highly capable, successfully detecting 23+ exoplanets and brown dwarfs to date~\cite{Wang2021_KPICPhaseI}. It has already enabled new science like: the first spin measurements of substellar companions like the HR-8799 planets~\cite{Wang2021_KPICScience,Morris2024arXiv240513125M}, exo-satellite searches around sub-stellar companions~\cite{Ruffio-DER2023,Horstman_RVM_2024}, characterization of hot-Jupiters~\cite{Finnerty2023AJ....166...31F,Finnerty2024AJ....167...43F}, and abundance measurements of directly imaged exoplanets and brown-dwarfs including isotopologues~\cite{Wang2021_KPICScience, wang-SDI2022, Wang2022_KPICCORetrieval,Xuan-ACV2022,Xuan2024ApJ...962...10X,Sappey2022_KPICHD206893,Costes2024A&A...686A.294C,DoO2024AJ....167..278D,Xuan2024ApJ...970...71X, Hsu2024ApJ...971....9H,Zhang2024arXiv240720952Z}.

KPIC phase II built on the success of phase I with goals of 1) improving planet throughput, 2) reducing stellar leakage, 3) accessing new spatial scales closer to the star, and 4) providing new observing modes/bands. A series of upgrades were carefully planned as part of phase II to address these requirements. The first upgrade was a 1000-element deformable mirror (DM), which allows for superior static aberration correction and speckle control at the location of the fiber. The second upgrade was a pupil mask mechanism supporting a microdot apodizer to suppress stellar leakage at intermediate separations~\cite{Calvin2021_KPICLab} and a vortex mask to enable vortex fiber nulling (VFN) at small separations~\cite{Ruane2018_VFN, Echeverri_VFN2023}. Third was a set of Phase Induced Amplitude Apodization (PIAA) beam-shaping lenses~\cite{Guyon2003_PIAA}, which match the shape of the telescope beam to the fundamental mode of the fibers, thereby boosting coupling and hence throughput~\cite{Calvin2021_KPICLab, Jovanovic2017_SMFOnSky}. In addition to these key upgrades, the PyWFS and tracking camera pickoffs were moved to mechanisms allowing for several light sharing options, and the fiber port was replaced with a stage that supports several lens/bundle combinations allowing KPIC to operate in additional wavebands. Phase II was built across 2020 and 2021 and underwent extensive laboratory validation in late 2021 and early 2022 before shipping to Keck for installation in February 2022. In April 2024, we performed a service mission which brought new dichroic options, lenses and a fiber bundle which operate across y-H band, an atmospheric dispersion corrector (ADC), and several laser frequency combs that were connected with KPIC. 

In this paper we provide an overview of the phase II instrument design and its laboratory characterization. Specifically, section~\ref{sec:design}, provides an overview of the phase II instrument layout. Section~\ref{sec:details} elaborates on the design providing great detail about the new modules/modes. This section is designed to be a resource and can be skipped if details aren't desired. Section~\ref{sec:lab_validation} summarizes the laboratory validation, while Section~\ref{sec:improved_cal} outlines enhanced procedures for compensating for static aberrations and non-common path errors. Finally, we present results from commissioning in Section~\ref{sec:commissioning} and highlight the performance of the final phase II version of the instrument. 

%%%%%%%%%%%%%%%%%%%%%%%%%%%%%%%%%
\section{Phase II Design Overview}
\label{sec:design}

%Key points to hit
%%% Walk through of phase II optical elements
%%% diagrams - schematics and photos
%%% Outline other upgrades to PyWFS, SAPHIRA, ZWFS etc. 

In this section, we provide an overview of the final as-built phase II instrument design after the 2024 service mission. The instrument leverages heavily from the requirements, design, and lessons learnt from the phase I instrument, so we refer the reader to a full description of the phase I instrument in Delorme~et~al.~2021~\cite{Delorme2021_KPIC}.

\begin{figure}[htbp]
\centering\includegraphics[width = \linewidth]{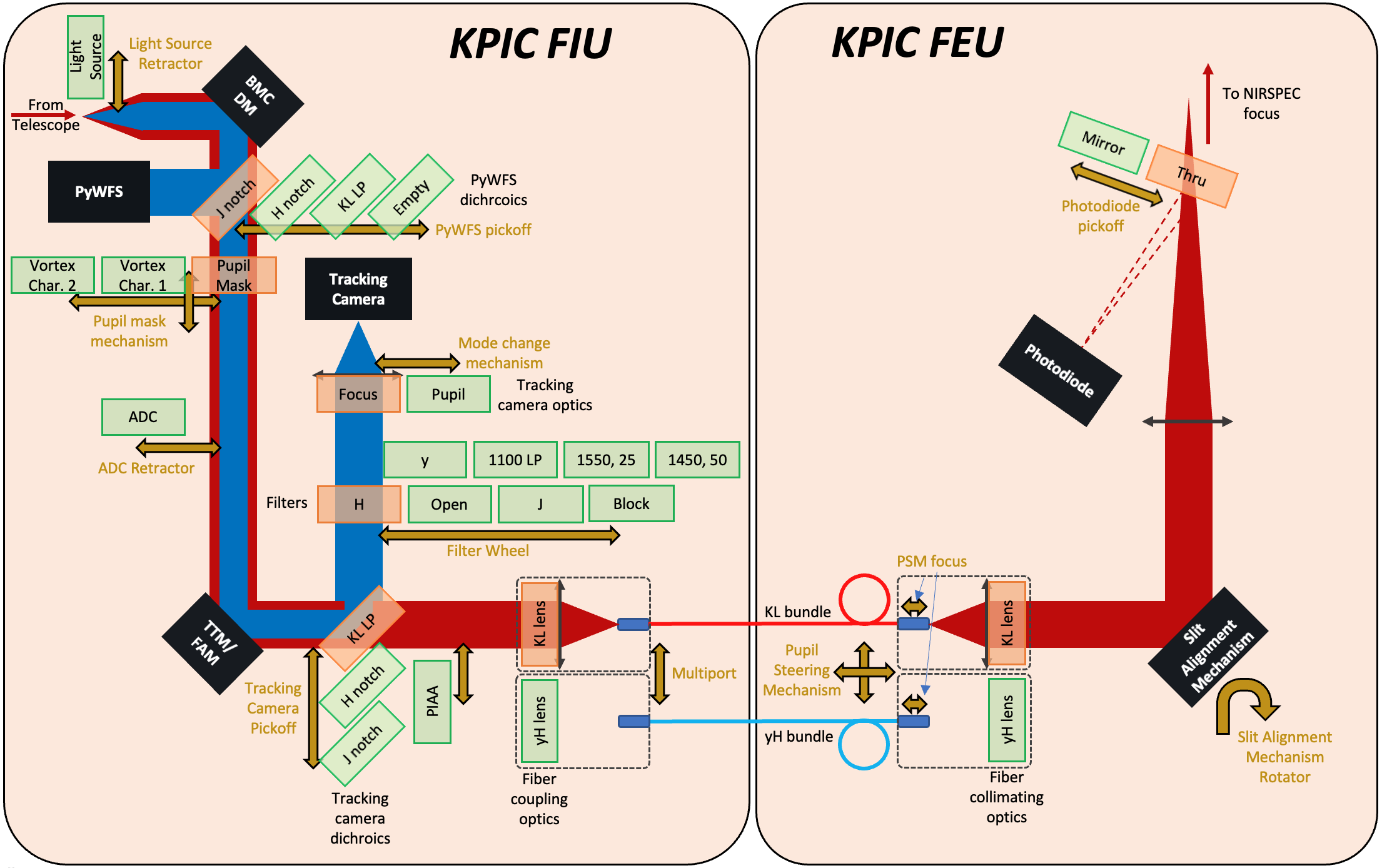}
\caption{A simplified diagram of the KPIC phase II instrument including the FIU and FEU. All moving mechanisms are denoted with double ended arrows. The options for each mechanism are listed next to them. The options highlighted in orange are the default settings used for nominal K band science operations most similar to the original phase I mode. The green options are alternative settings available for the many phase II modes. \label{fig:schematic}}
\end{figure}

\begin{figure}[htbp]
\centering\includegraphics[width = \linewidth]{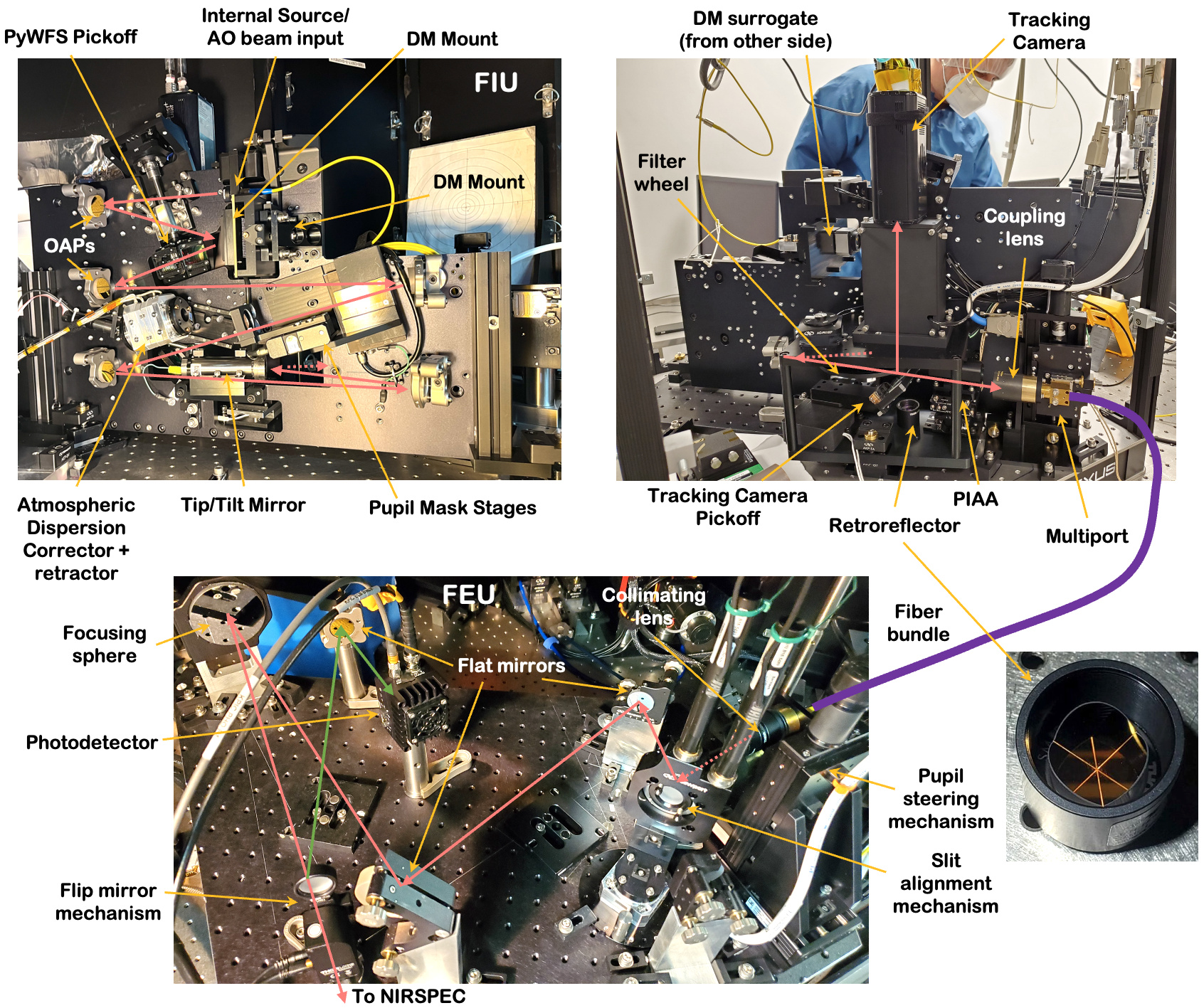}
\caption{(Top) Photos of the phase II version of the FIU. (Left) Front side of the instrument taken during the service mission in April 2024. (Right) Rear side of the instrument taken in December 2022. (Bottom) A photo of the new FEU taken at Keck inside the Keck AO bench. Inset, image of the retro-reflector used in  the FIU. \label{fig:instrument}}
\end{figure}

A schematic diagram of the KPIC beam train is shown in Fig.~\ref{fig:schematic}. At a top level, the diagram is split into the Fiber Injection Unit (FIU) on the left and the Fiber Extraction Unit (FEU) on the right. The FIU's primary purpose is to efficiently inject AO-corrected light into the SMFs of the fiber bundle. The FEU re-images and interfaces the output of the bundle to the existing NIRSPEC optics, which were originally designed for slit-based operations rather than fiber-based. Figure~\ref{fig:instrument} provides front and back photos of the FIU plate, as well as a photo of the FEU. %A photo of the phase II FIU plate as well as the FEU is shown in Fig.~\ref{fig:instrument}. 
The design of phase II is compatible with that of phase I so that the old plate could be easily replaced with the phase II variant: we used a similar optical layout (beam heights off the plate and bench), the same off-axis parabola (OAP) prescriptions and relays, the same vertical layout, and kinematic features enabling a direct swap of the two plates at the observatory. Unlike the phase I plate, the phase II plate and opto-mechanics were black-anodized to minimize stray light scattering and unwanted reflections.  
%For a detailed description of the Phase II optical design we refer the reader to our previous work~\cite{Jovanovic2020_KPICPhaseII}.

%The classical HDC observing mode supported in Phase I where light from a known companion is directed onto, and injected into, the fibers in the bundle for spectral characterization. This mode offers a fiber for the science target, at least one fiber for the speckle field to get a contemporaneous spectrum of the host, and a background fiber. When observing in the K or L bands, we usually bounce the target between two fibers during the observation to calibrate systematics due to the background and fringing from the window in NIRSPEC. This mode works well for targets at $\gtrsim 5\lambda/D$ from the optical axis. In Phase I, we successfully offset to a companion as wide as 3.6 arcseconds ($80\lambda/D$). 

\subsection{The Fiber Injection Unit}
This section follows the instrument beampath within the FIU, starting at the input focal plane and ending at the input to the science fiber bundle. The top left panel in Fig.~\ref{fig:instrument} shows the instrument as viewed from the front side. In the image, the input focal plane is occupied by the internal source fiber as labeled, but this source can be retracted from the optical axis to allow the beam delivered by Keck AO to enter the system. When inserted, the source fiber allows the system to be internally calibrated independent of Keck AO, which was determined to be a valuable additional capability from phase I. 

The injected beam is collimated by an OAP and the resulting pupil is relayed onto the 1000-actuator DM (BMC DM; Boston Micromachines Kilo-DM). Prior to the DM's installation, a flat mirror was temporarily mounted in a similar way to how the DM is mounted so it could be rapidly swapped to simplify initial alignment (see right panel of Fig.~\ref{fig:instrument}). The beam then passes to the PyWFS pickoff, which holds dichroics that select the wavelengths to be used for wavefront control by the PyWFS. The original phase I pickoff was static but was upgraded in phase II to a rotating mechanism up to 4 dichroics to provide additional options. The NIR PyWFS was used throughout phase I and briefly at the start of phase II but was decommissioned in January 2024 after a drop in detector QE built upon other issues. Wavefront control during KPIC observations is now provided exclusively by the Shack-Hartmann wavefront sensor (SHWFS)~\cite{Wizinowich2000}. However, the dichroics in the PyWFS pickoff mechanism are still utilized for KPIC observing because they shift the pupil laterally to center it on the downstream OAPs as originally designed, thereby minimizing aberrations in the system.

The beam transmitted through the PyWFS dichroics is then re-imaged by an OAP relay to a downstream pupil plane where the pupil mask mechanism is located. This mechanism has 3 options: a through-port that passes the beam without modification, and two slots for masks. When we deployed phase II in 2022, we installed a microdot apodizer~\cite{calvin_EDE2021} and a charge 2 vortex in the two mask slots. However, it was determined through on-sky measurements that the microdot apodizer offered no benefits on-sky (see Sec.~\ref{sec:MDA}), so it was swapped for a charge 1 vortex mask in the 2024 service mission (see Sec.~\ref{sec:VFNDesign}). This is an alternative use of a vortex mask, best suited to look for companions located at $0.5-2~\lambda/D$ next to the host star (See Sec.~\ref{sec:VFNDesign}). The mechanism supports these modes by translating the chosen mask into, and out of, the beam and allowing for fine positioning to align it with the beam. Unlike conventional coronagraph implementations, KPIC chose to locate these optics in a pupil plane, simplifying the system and leveraging the fact that pupil-plane VFN uses the same F/\# as that used for normal fiber injection into an SMF\cite{Ruane2019SPIE}. After the pupil mask mechanism is an ADC which corrects for chromatic smearing in the point spread function to maximize broadband coupling and minimize stellar leakage in the nulling mode (see Sec.~\ref{sec:ADCDesign}). The ADC optics are mounted on a carriage that allows them to be retracted from the beampath whenever they are not needed. This is important for minimizing transmission losses through the glass prisms when atmospheric dispersion is not a limiting term. After the ADC, another OAP relay is used to re-image the pupil plane onto the tip/tilt mirror (TTM, Physik Instrumente, S-330.8SL with an E-727.4SD controller). The TTM is used to move the target around the fiber focal plane as needed to align it with the optical fiber of choice and is therefore also referred to as the Fiber Alignment Mirror (FAM). The piezo based TT stage has hysteresis, but this is eliminated when we operate in closed loop using the built-in strain gauge sensors which is the default operating state for KPIC. The stage combined with the tracking camera has a capture range of $\pm$2.35 arcsec in closed loop. The stage has a repeatability of 0.35 mas or $1/107^{th}$ of a $\lambda/D$ at 2 $\mu$ms - D is
the telescope diameter and $\lambda/D$ is the angular resolution of the telescope. Because of this, the stage can be reliably returned to a given position when trying to maximize coupling for example. 

Due to space constraints within the Keck AO bench, KPIC uses both sides of the vertical plate. The right panel in Fig.~\ref{fig:instrument} shows the rear side of the instrument with the TTM also serving to send light from one side to the other. The reflected beam is first incident on a flat mirror which sends the beam on its final trajectory towards the fiber focal plane. Before it reaches the fibers, the beam meets the tracking camera pickoff (TCP) which has also been upgraded in phase II to a switchable, translating mechanism supporting up to three dichroics in phase II. The light reflected by the selected pickoff dichroic is sent to the tracking camera which is used to visualize and determine the location of the target and drive the TTM to put the target on the fibers as desired. On the way to the tracking camera, a custom filter wheel supports many options for controlling the passband used for tracking or for blocking the light entirely. As in phase I, the tracking camera retains the ability to switch between focal plane viewing, used for tracking, and pupil plane viewing, used for aligning the pupil masks. In addition, a Zernike mask is installed near the intermittent focal plane in the tracking camera optics box and enables Zernike wavefront sensing\cite{NDiaye2013_ZWFS, VanKooten2022_KPICZWFS}. 

The beam transmitted through the tracking camera pickoff is next incident on the PIAA lens assembly. This consists of two pre-aligned aspheric lenses in a tube which remap the pupil illumination from an obstructed flat top to a soft edged (quasi-Gaussian) beam that better matches the fundamental mode of the SMF. For more details about the PIAA optics, see Sec.~\ref{sec:PIAADesign}. The PIAA optics are on a translation stage, allowing them to be inserted and retracted as needed. After the PIAA, the final FIU mechanism is called the ``multiport" which is a stage supporting up to three coupling lenses and three corresponding bundles. This replaces the static coupling OAP used in phase I with high performance lenses. Lenses were chosen in phase II due to the performance limitations of using a fast OAP across a wide field. Additionally, it is much easier to mechanically parallelize multiple bundles if each one has its own set of injection lenses. For the KL bundle, a custom triplet lens was designed with a focal length of 37~mm and consisted of CaF$_{2}$, ZnS, and AMTIR-1 lenses. The triplet was designed to have extremely low wavefront error out to 2 arcseconds so it can efficiently couple into any of the KL bundle fibers and to be achromatic from K-L bands. The measured wavefront error across that field of view was validated to be $<$32~nm RMS. It was manufactured by Rainbow Research Optics and packaged by Photon Gear. Leveraging the upgraded multiport design from 2022, the 2024 service mission brought a new lens and bundle optimized for science in y, J and H bands. This is explained in more detail in Sec.~\ref{sec:yHBundleDesign}. Users can therefore now execute science observations in y, J, H, K, or L~band by selecting between the KL and y-H bundles and choosing the appropriate upstream dichroics.

KPIC phase I utilized a retro-reflector located beneath the tracking camera pickoff to steer retro-fed beacons from the bundle to the tracking camera so the location of the fibers in the bundle could be determined. However, due to the high wavefront error in the retro-reflector used in phase I (a Thorlabs PS975M-C), the beacons were aberrated and defocused making them challenging to use. In the 2024 service mission, a new all-reflective design was deployed (see inset in the lower right of Fig~\ref{fig:instrument}). It provides superior reflected wavefront error ($<$30~nm RMS across an 18~mm aperture) with $<$5~arcsec beam deviation (vendor was PLX). These performance levels addressed the limitations of the previous device so that sharp retro-reflected images are now obtained.

\subsection{The Fiber Extraction Unit}
\label{sec:details}
%New FEU --\> PD, detached bundle from NIRSPEC --\> faster and easier calibrations and also improved maintainability since no servicing needed when NIRSPEC goes in/out.

%outline new stop - 8.1 mm down to 5 mm, reduced thermal background

Beyond the upgrades to the FIU, phase II also overhauled the FEU to improve the speed and quality of the instrument calibrations and to make the interface between KPIC and NIRSPEC more robust and reliable. 

For KPIC observations, NIRSPEC is moved into the Keck AO room and parked in front of the AO bench. In phase I, KPIC required a telescope crew member at the summit to manually connect the FEU-end of the fiber bundle to the FEU optomechanics installed inside the calibration unit of NIRPSEC. This increased the possibility that the fiber could be damaged either during the connection procedure or while it hung between NIRSPEC and the AO bench during observation. To eliminate the need to manually connect and disconnect the fiber for each run, and to reduce the likelihood of damaging the fibers, a new FEU was built inside the Keck AO bench itself. This new FEU is located on the NIRSPAO plate, which is an optical relay that normally allows for the Keck AO beam to be sent directly to NIRSPEC without fibers in what is known as NIRSPAO mode. We installed the FEU such that we can inject the output of the bundles into the NIRSPAO optics and project the image onto the slit of NIRSPEC. Specifically we use a mechanism very similar to the multiport to recollimate the output of the bundles. This is called the Pupil Steering Mechanism (PSM) because it uses two translation stages to translate the lens/bundle combination which effectively translates the pupil. This can be used to center the exit pupil from the fibers onto the cold stop inside NIRSPEC. The pupil is first projected onto a steering mirror, referred to as the slit alignment mechanism, which steers the beam across the slit of NIRSPEC. The steering mirror is mounted on a rotation stage which allows it to be removed and re-inserted into the beam to allow the traditional NIRSPAO mode to be maintained. The beam then bounces off 3 flat mirrors and a spherical mirror which focuses the beam. A pickoff mirror is located at the exit port of the Keck AO bench and can be flipped into, and out of, the beam to send the light to a photodetector (PD) for daytime calibrations. These FEU upgrades have greatly simplified and sped up pre-observing preparations, improved the resulting calibration quality, and mitigated a key risk to the bundle recognized in phase I. See Sec.~\ref{sec:newCals} for additional details.

\section{Design Details}
%\subsection{Upgrade details}
%plot a figure with calibration possibilities and describe LFCs. Could include LFC spectra too. 

The instrument design presented in the previous section is the result of several individual upgrades to sub-modules within the instrument in phase II. Taken all together, these upgrades enhance KPIC's capabilities by increasing the instrument throughput, unlocking new observing modes, and extending science wavelength coverage. This section covers the individual upgrades in more detail, explaining the motivations for each, how the upgrade was executed, and how it improves the instrument's capabilities. This section serves as a resource for the reader interested in further details on any aspect of the design. Skip this section if details are not desired.

\subsection{New dichroics and filters}
One of the key goals for phase II was to expand the operating wavelength range beyond the original K and L bands to include the y, J, and H bands. This required that new dichroics be installed. However, phase I operations revealed that the highly parallel dichroics originally used in the instrument introduced time-varying fringing in the NIRSPEC spectrum, which proved difficult to calibrate out and was one of the key limitations in phase I~\cite{Finnerty_CHJ2023,Katelyn_FAF2024}. To mitigate this effect, new dichroics were developed on wedged substrates such that all dichroics in phase II have a nominal wedge angle of $50\pm10$ arcsecs. This ensures that the co-propagating ghost is offset by at least 5$\lambda/D$ ($D$ is the telescope diameter and $\lambda/D$ is the angular resolution of the telescope)  at the longest operating band for each dichroic. This offset leads to a sufficiently high contrast between the primary science beam and the ghost to minimize the visibility of the fringes.   

\begin{figure}[htbp]
\centering\includegraphics[width = \linewidth]{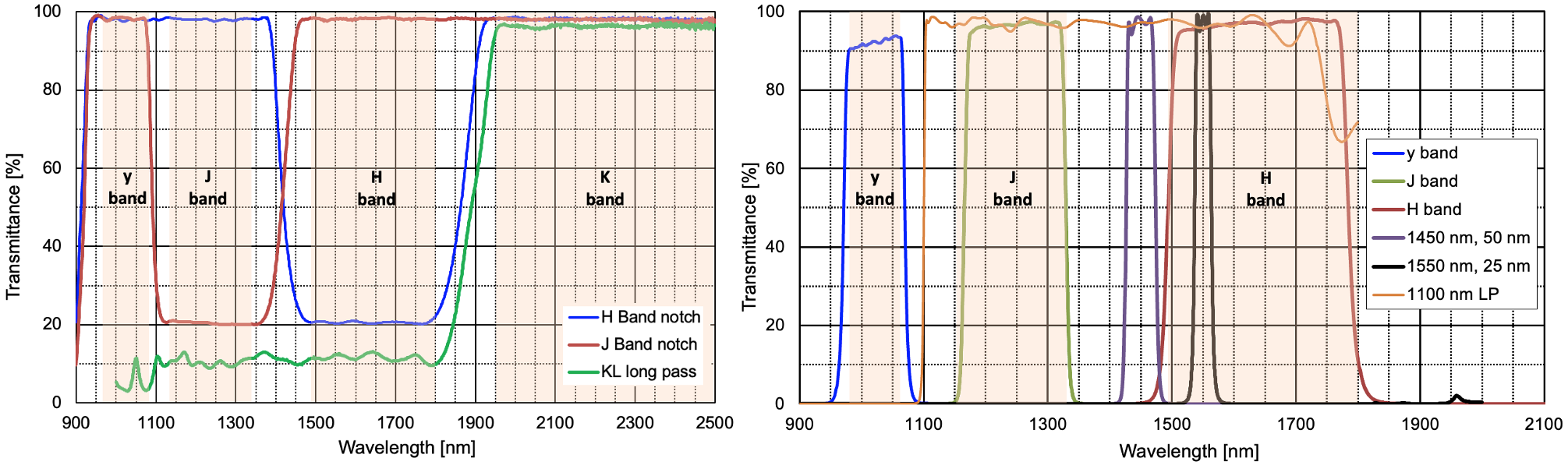}
\caption{The spectral profiles of (Left) the dichroics and (Right) the filters in the FIU. The KL long pass dichroic extends out to the end of L band, but we only show spectra out to K band so the details of the shorter wavelength region are more visible. \label{fig:dichroicsfilters}}
\end{figure}

The selection of dichroics installed in both the PyWFS pickoff and tracking camera pickoff mechanisms are listed in Fig.~\ref{fig:schematic} and their transmission is shown in the left panel of Fig.~\ref{fig:dichroicsfilters}. The J and H notch dichroics are on fused silica substrates. The KL long pass dichroics were deposited on a CaF$_{2}$ substrate to ensure high transmission into L band. The transmitted wavefront error across the beam is ${<}10$~nm RMS, while the reflected wavefront error is ${<}22$~nm RMS for all dichroics. These levels of wavefront error are ideal for minimizing non-common path aberrations between the fiber and tracking camera focal planes and minimizing the overall static aberrations to facilitate high coupling efficiencies. 

For completeness, the filters in the tracking camera beam path are listed in Fig.~\ref{fig:schematic} and are shown in the right panel of Fig.~\ref{fig:dichroicsfilters}. Given all the available permutations, Table~\ref{tab:DichroicSettings} lists the dichroic and tracking filter wheel settings now used for each science band, along with the resulting throughput in the tracking band. This throughput defines the flux available to the tracking system and hence sets the limiting guide star magnitude. In phase I, K band science used the same settings as the current L band in the table, meaning that tracking was limited to 9\% throughput. As such, the phase II upgrade also increased the tracking flux by $10{\times}$ for K band science. Note that the reported throughput in the Table is only the contribution from the dichroics and filter optics; there are additional losses from other optics in the beam path and from the atmosphere.

\begin{table}[]
\begin{center}
\begin{tabular}{|c|c|c|c|c|c|}
    \hline
    \begin{tabular}[c]{@{}c@{}}Science\\ Band\end{tabular} & 
    \begin{tabular}[c]{@{}c@{}}PyWFS Pickoff\\ Dichroic\end{tabular} & 
    \begin{tabular}[c]{@{}c@{}}TCP \\ Dichroic\end{tabular} & 
    \begin{tabular}[c]{@{}c@{}}Filter \\ Wheel\end{tabular} &  
    \begin{tabular}[c]{@{}c@{}}Tracking \\ Band\end{tabular} &
    \begin{tabular}[c]{@{}c@{}}Tracking \\ Throughput (\%)\end{tabular} \\      \hline
    y  &  J-notch       &  H-notch       &  H Filter  &  H band  &  80  \\      \hline
    J  &  H-notch       &  H-notch       &  H Filter  &  H band  &  16  \\      \hline
    H  &  J-notch       &  J-notch       &  J Filter  &  J band  &  16  \\      \hline
    K  &  J-notch       &  KL long pass  &  H Filter  &  H band  &  90  \\      \hline
    L  &  KL long pass  &  KL long pass  &  H Filter  &  H band  &  9   \\      \hline
\end{tabular}
\caption{
    \label{tab:DichroicSettings}Dichroic and filter settings for the tracking arm when observing in the given KPIC science bands. We note that this is only the throughput/transmission due to the three optics in columns 2-4; there are additional losses from the atmosphere and other optics in the beam path. We also note that 100\% of the science band is passed to the fiber bundle, when considering this metric of just the throughput due to these three optics. TCP - tracking camera pickoff. 
    }
\end{center}
\end{table}

\subsection{y-H optics and bundle} \label{sec:yHBundleDesign}
\begin{figure}[htbp]
\centering\includegraphics[width = \linewidth]{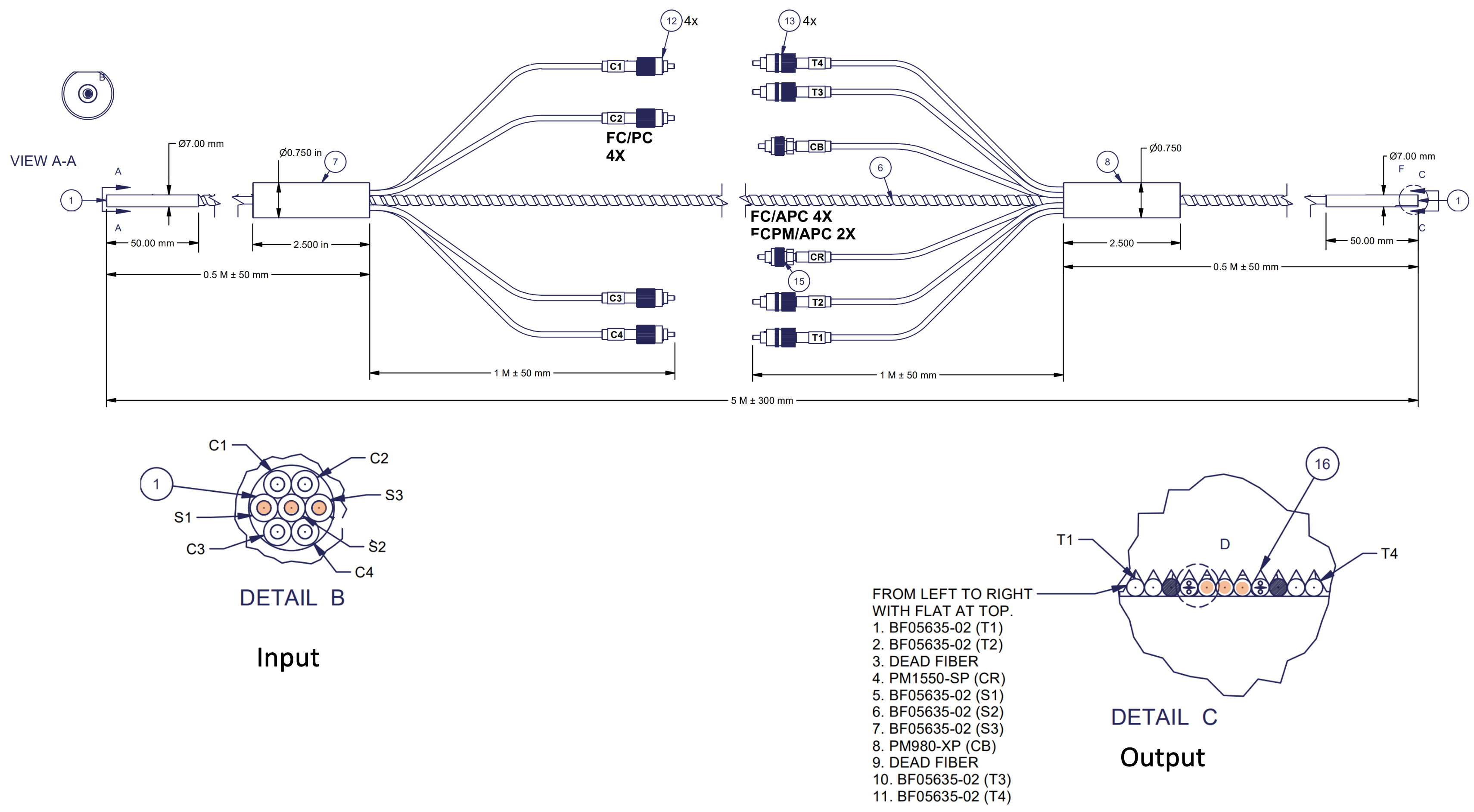}
\caption{The CAD drawing of the yH fiber bundle. Call out 1 indicates the 7 mm diameter input and output connectors. Call out 7 and 8 indicate the breakout junctions. Call out 12 shows the FC/PC connectors used on the breakouts at the input. Call out 13 shows the FC/APC connectors while 15 highlights the PM FC/APC connectors. Call out 16 indicates the vgroove used at the output. S1-3 highlight the science fibers also identified by the orange circles. C1-C4 are the reverse injection pigtails. T1-T4 could be used to determine the location of the bundle on the NIRSPEC slit viewing camera. CR and CB are the calibration channels designated for the laser frequency combs. All connector ends were polished and are free from anti-reflection coatings.   \label{fig:bundle}}
\end{figure}

Enabling y-H science also required a new fiber bundle and new coupling and collimating lenses. A CAD drawing of the bundle fabricated by Molex/Fiberguide Industries is shown in Fig.~\ref{fig:bundle}. 
The fiber bundle shares many similar attributes to the existing KL bundle described in Delorme~et~al.~2021~\cite{Delorme2021_KPIC}, including: 7~mm diameter round connector ends, a 5~m length, science fibers routed from the input to the output connectors, pigtails provided at both ends so that light can either be reverse injected into the FIU or sent directly to the FEU/NIRSPEC bypassing the FIU, and a v-groove at the output to align the fibers with the slit of NIRSPEC. The main differences are that the new bundle utilizes silica optical fibers, due to the shorter wavelength operation, and that it provides only 3 sky viewing fibers (highlighted with orange circles in the figure and labelled S1-S3). The fiber used was OFS SMBD0980B, a fiber with a single-mode cutoff around 980 nm (the short end of y-band). This fiber was selected because the IRD instrument~\cite{kotani2018} at Subaru Telescope validated low bend loss performance through to the end of H band. The figure shows additional fibers on the input end labelled C1-C4 which are used to reverse inject light into the FIU to determine the location of the bundle on the tracking camera. On the output end, fibers T1-T4 were designed to be used to spot the bundle and determine its orientation on the slit viewing camera in NIRSPEC. Given that NIRSPEC uses slits that have a reflective outer region, light from outside the slit can be steered to the slit viewing camera. S1-3, C1-4 and T1-4 are all based on the OFS fiber described above. Two additional calibration fibers labelled CB (calibration blue) and CR (calibration red) are provided at the output of the bundle. These fibers are meant to carry the light from the laser frequency comb and are based on polarization maintaining (PM) fibers to fix the polarization of the combs with respect to the diffraction grating (PM-980-XP for CB and PM1550-SP for CR). This allows KPIC to have simultaneous comb light and science light on NIRSPEC during observations. The magnification of the FEU is such that only 4 fibers can be passed through the NIRSPEC slit at once, which means the user can choose from CR or CB for the simultaneous calibration trace, but not both at once. All connector ends of the bundle were polished and left uncoated. All connectors on this bundle were cleaned with lens tissue soaked in isopropyl alcohol. We twist the tissue on the surface of the fibers and drag to remove the excess liquid. We then inspect the fiber ends under a microscope (stationary or handheld). Once the bundle were installed at the summit, the fiber ends are never handled or cleaned again. This has proven sufficient in KPIC phase I. 

The bundle was characterized in the laboratory. Note that the measured throughput (i.e. power after the bundle/power before the bundle), includes the coupling efficiency to the bundle, Fresnel reflections and the fiber transmission (which is 100\%). The maximum theoretical coupling efficiency for the beam at Keck to the bundle would be 62.1\% at 1050 nm, 59.1\% at 1310 nm and 56.6\% at 1550 nm (imposed by the overlap between the centrally obscured Airy beam and the fiber mode). The measured throughput, normalized by the theoretical coupling efficiency so we can determine the bundle efficiency in isolation, was determined and is summarized in Table~\ref{tab:bundlethroughput}. It can be seen that S2, the central fiber has the highest efficiency (80-90\%), S3 the second highest (65-77\%) and S1 the lowest (46-50\%). These results were double checked and the different efficiencies for various ports are indeed real. The throughput also increases slightly as the wavelength increases. Finally, the absolute throughputs include the Fresnel reflection losses incurred at both ends of the bundle which contributes $\sim$8\% to the efficiency reduction. To clarify, we had many issues with these bundles and had to work with the vendor to iterate several times. Through this process, we ran out of time to coat the bundle and we only identified the non-uniform efficiency shortly before deployment. However, at shorter wavelengths the thermal background is much lower, so bouncing between fiber cores which is typically carried out during K band observations to remove the background is not necessary. As such, a single high-efficiency channel is sufficient for the shorter wavelengths so S2 is the baseline for science going forward.   

\begin{table}[t]
\begin{center}
\begin{tabular}{|c|c|c|c| }
\hline
& \multicolumn{3}{|c|}{Throughput (\%)} \\
\hline
Fiber \# & 1050 nm & 1310 nm & 1550 nm \\
\hline
\hline
S1 & 45.7 & 49.3 & 50.4 \\ 
S2 & 80.2 & 84.8 & 89.4 \\ 
S3 & 64.5 & 75.5 & 77.1 \\ 
\hline
\end{tabular}
\caption{Throughput of yH fiber bundle measured in the laboratory}
\label{tab:bundlethroughput}
\end{center}
\end{table}

To efficiently couple into the yH optical bundle, a custom doublet lens was designed. Given the mode field diameter properties of the optical fiber, and the pupil diameter in KPIC (12.6~mm), a lens with a custom focal length of 47~mm was designed for optimal coupling. The custom lens is a bonded doublet utilizing CaF$_{2}$ and F5. The doublet is necessary to minimize chromatic defocus and maximize coupling to the optical fibers over the large bandwidth and the finite field size. The Strehl of the lens was measured by the vendor (SALVO technologies) and it was determined to be 99.5\% on-axis and $>$98.5\% over the field at a wavelength of 1~$\mu$m. The measured throughput through the anti-reflection coated lens was $\sim$99\% across 1050, 1310 and 1550 nm. The lens was installed on the multiport with the yH bundle. At the output, a second identical lens is used to collimate the beam to feed it into the FEU and eventually into NIRSPEC.

\subsection{Improved Calibration Infrastructure} \label{sec:newCals}
\begin{figure}[htbp]
\centering\includegraphics[width = \linewidth]{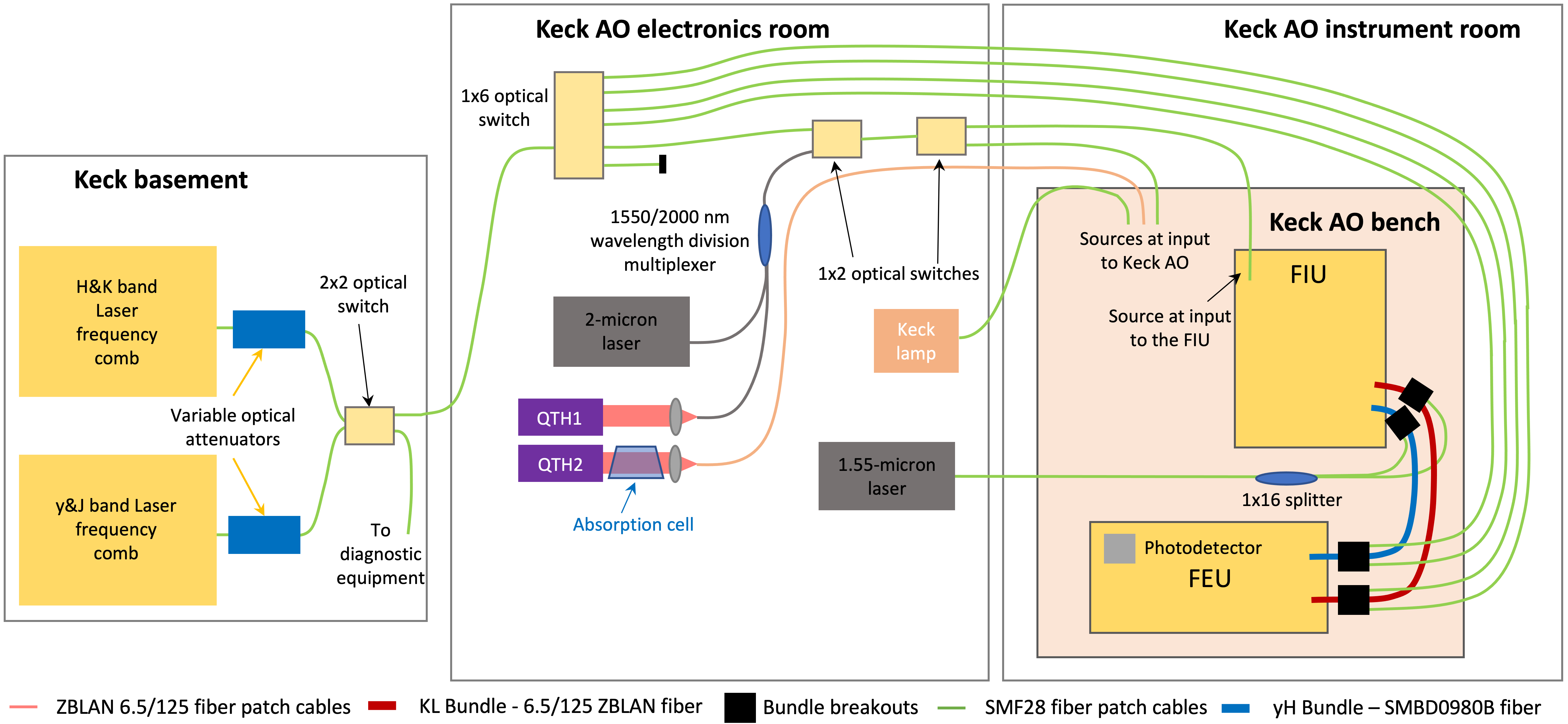}
\caption{A schematic diagram of the calibration scheme for KPIC. MEMS-based optical switches are used to route the various calibration sources to either the input of the Keck AO bench, the input to the FIU, or the breakouts on the input of output of the two fiber bundles. QTH stands for Quartz Tungsten Halogen. The schematic does not show any of the optics, just the fibers and locations where the light sources can be routed to. The schematic also does not show the fiber optic connections.   \label{fig:calibration}}
\end{figure}

The KPIC instrument also received a series of upgrades to improve the ability to calibrate the instrument. During daytime calibrations two main routines are used to optimize the injection: fiber finding which involves scanning the TTM across the fiber core to determine the optimum alignment, and a static aberration compensation routine which relies on scanning the amplitude of Zernike modes applied to the DM to compensate for aberrations and improve coupling efficiency. The key to efficiently executing extensive scans in a reasonable time frame is to use a sufficiently bright light source so that a fast photodetector can be used. This is because the NIRSPEC detectors have slow read out times. For calibration of the KL bundle, a quartz tungsten halogen (QTH1, Thorlabs, SLS202L) lamp is multiplexed with a 2-micron laser source via a wavelength division multiplexer (WDM, Thorlabs, WD1520FB) as seen in Fig.~\ref{fig:calibration}. This provides a bright 2-micron light source in the science bands which can be used for optimizing coupling with a PD (PDA10DT). Owing to its broadband nature, the QTH provides a beacon that can be seen by the C-red2 tracking camera. This is critical as the fiber finding technique does not rely on the absolute knowledge of the TTM angle but rather the sub-pixel position on the tracking camera to determine where maximum coupling occurs. Those light sources can be directed to the input of the Keck AO bench or the input of the FIU via a series of 1$\times$2 optical switches (Agiltron, FFSM-122C01333). It is advantageous to inject light at the front of the AO bench so that all static aberrations in the Keck AO+KPIC beam trains can be compensated. 

Because the WDM is based on silica fibers, it does not transmit at $>2.2~\mu$m. In addition, the PD is not sensitive above $2.5~\mu$m. Therefore, for L band daytime calibrations, the 2-micron laser+QTH combination is used to find the fiber and compensate for aberrations in K band, but a second QTH injected into the Keck AO bench is used to prepare the spectrometer. This source has much higher L band signal owing to the ZBLAN fiber used to transport the light to the input of Keck AO. To switch fibers, a translation stage is used to move the desired fiber into the focus of the telescope. A single 10 cm long absorption cell with a mixture of methane ($\sim23$~Torr), nitrous oxide (38 Torr), acetylene (38 Torr) and carbon dioxide (106 Torr) (fabricated by Wavelength References) is located between the second QTH (QTH2) and a fiber coupling lens which imprints a spectrum that can be used for spectral calibration. These gases and partial pressures were carefully chosen to maximize the number and depth of spectral features across the K and L bands, while preventing the spectral features from being resolved in NIRSPEC. For completeness, there is another source provided more generally to the Keck AO bench named the ``Keck lamp", which is sometimes used by KPIC, but is mostly used by other instruments. 

To determine the location of the fibers on the tracking camera initially, a 1.55~$\mu$m laser diode is used. Light from the source is split by a 1$\times$16 fiber splitter (Thorlabs, TDS1315HF) and routed to the breakouts on the input end of both bundles. 6 ports are needed for the KL bundle and 4 for the yH. When the laser is turned on, beams are broadcast in reverse through the FIU, which bounce off the back off the tracking camera dichroics, and are steered back up via a retro-reflector, which partially pass through the dichroic the second time and broadcast beacons on the tracking camera. This mode was critical to centering both bundles on the tracking camera initially. It can also used to track any drifts between the tracking camera and fiber focal planes.  

For spectral calibration, KPIC phase II benefits from two laser frequency combs (LFCs). LFC calibration light in the y and J bands was made available from the Keck Planet Finder (KPF) instrument. This mode-locked LFC, manufactured by Menlo Systems~\cite{Probst_ALF_2014}, is used by KPF to calibrate their precision radial velocity system from 400-840 nm, but also produces near-infrared light. It provides calibration lines referenced to a GPS-disciplined rubidium clock at 20 GHz line spacing. In collaboration with the KPF team, we modified the comb by installing a near infrared pickoff to make use of the unused light from 900-1500 nm. As is common with LFCs used in astronomy, a custom optical spectral flattener was built by Menlo using liquid crystal on silicon (LCOS) spatial light modulator (SLM) technology to provide uniform intensity across the bandpass, and installed in the IR arm of the KPF LFC to flatten the spectrum for KPIC. The output light is then passed through a variable optical attenuator (VOA) (Agiltron, DD-100-11-980/1310-6/125-P-50-3A3A-1-1-485:1-6-MC/RS232) which allows for the brightness to be adjusted. A fiber shutter (Agiltron, FFST-211141323) was installed past the attenuator. 

The second LFC is a custom-built electro-optic (EO) modulation frequency comb similar to that first demonstrated at the Keck Observatory~\cite{Yi_2016}, the EO comb installed at the Hobby Eberly Telescope for the Habitable Planet Finder (HPF) instrument~\cite{Metcalf}, and the infrared  laser comb at the Subaru Observatory~\cite{Subaru}. It provides 16 GHz-spaced calibration lines with the spacing locked to the same model of GPS-disciplined rubidium clock as that used by the Menlo comb (SRS FS725). The pump laser is line referenced to a separate rubidium standard (Vescent D2-210-Rb). It has been optimized for H and K band operation, but also produces light from 1200 nm to red of 2200 nm. Further details of the H-K LFC design can be found elsewhere~\cite{Leifer_inprep}. Another Menlo spectral flattener was installed for this LFC. The output of the flattener is similarly passed through a VOA and a shutter. Some of the light from the H-K LFC pump line at 1560 nm is tapped off, sent through a separate VOA and overlaid with the rest of the H-K LFC light to provide a wavelength flag on the spectrograph detector.

A 2$\times$2 MEMS-based switch (Agiltron, FFSW-222C00323) is used to select which LFC to send from the basement up to the Nasmyth platform where the AO room is located. After the 2$\times$2 switch, light from either LFC is transported from the basement to the Nasmyth by a previously-installed SMF28 fiber. This 60+m line of silica fiber greatly attenuates the K band spectrum that should be available from the second LFC. The HISPEC instrument, a new y-K band spectrometer with a dedicated fiber injection unit is set to arrive at Keck in 2026 and will replace these fibers with higher transmission ZBLAN fibers that will pass the K band light more efficiently~\cite{Mawet_FFH2024}. In the AO electronics room, a 1$\times$6 MEMS-based (Agiltron, LBSA-0065S10323) optical switch is used to further direct the light. The LFC signal can be routed to the breakouts at the output ends of either fiber bundle to be used as a simultaneous calibration trace while science is being conducted. They can also be steered to the front of the AO bench or the FIU so that the entire beam train can be calibrated if needed.

\subsection{NIRSPEC upgrades}
\begin{figure}[htbp]
\centering\includegraphics[width = \linewidth]{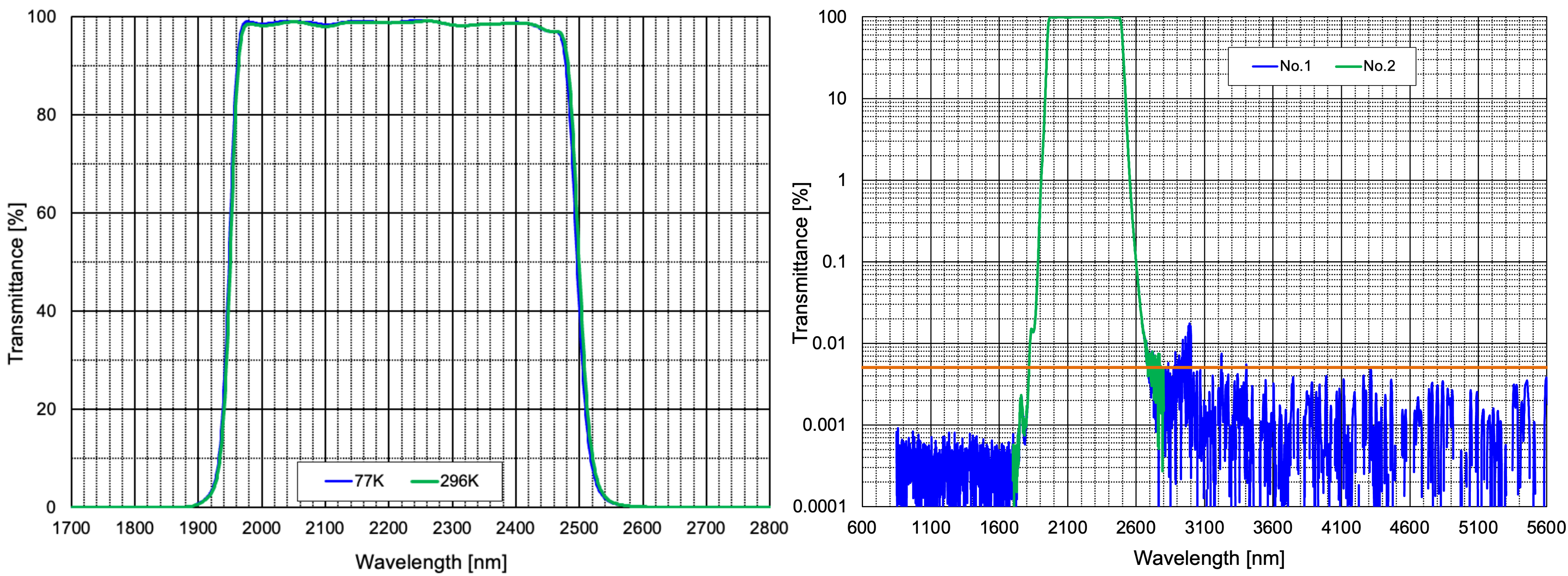}
\caption{The spectral profile of the new K band filter for NIRSPEC. (Left) Transmission profile on a linear scale. (Right) Transmission profile over a large bandwidth on a logarithmic scale to show the out of band suppression. The orange line is the requirement for suppression. \label{fig:Kband_filter}}
\end{figure}

Several optics in NIRSPEC were upgraded in March 2022. Firstly, the K short filter used in phase I observations was upgraded to a new filter produced by Asahi-Spectra. This filter utilizes IR fused silica and its transmission profile extends up to 2.50~$\mu$m to transmit the CO band head (see left panel in Fig.~\ref{fig:Kband_filter}). Compared to the previous filter, the new filter has $>50\%$ more transmission ($\sim$98.8\% average transmission efficiency from 1970-2480~nm), and much deeper suppression out-of-band (see right panel in Fig.~\ref{fig:Kband_filter}). The requirement for out-of-band suppression was 5$\times$10$^{-5}$ as indicated by the orange line, and the plot shows that this was met across nearly the entire spectrum. Subsequent testing with NIRSPEC validated the suppression was within specification. For completeness, the glass substrate used for the filter had a transmitted wavefront error of $<$7~nm RMS prior to coating and is unlikely to have changed post coating, and parallelism of $\sim2$~arcsec. 

The new FEU required a different collimator focal length to project the beam onto the slit alignment mechanism in the pupil plane. This changed the size of the beam imaged onto the cold stop inside NIRSPEC. To minimize thermal leakage, the cold stop in NIRSPEC was also updated from the 8.1~mm version used with the old FEU, to a 4.9~mm one with the new FEU.  

Finally, after a separate routine maintenance of NIRSPEC it was determined that the background levels had increased up to a factor two from some thermal light leak. A study was conducted to determine what the cause was but none could be found. This impacts KPIC's faint-object sensitivity by up to a $\Delta$mag = 0.4 \cite{Wang2024_KPICContrastCurve}.  

\subsection{Vortex Fiber Nulling} \label{sec:VFNDesign}
Since the concept of vortex fiber nulling was introduced, it was clear that the relatively simple mode would be a good fit for KPIC and could open a new parameter space for the instrument. Pushing for even smaller separations, this mode is optimized for the detection and characterization of targets between 0.5 and 2.0 $\lambda/D$ ($20-100$~mas in K-band). An optical vortex mask is used to imprint a spiral phase ramp onto the beam, preventing on-axis light from coupling into the SMF. Off-axis objects within an annular ``donut" profile, however, couple in and are transmitted to the spectrometer~\cite{Ruane2018_VFN,Echeverri2019b_VFN}. Due to the azimuthal symmetry of the coupling region, previously-unknown companions can be detected. VFN also enables spectral characterization of objects at spatial scales inaccessible to conventional coronagraphs. For VFN-mode observations, the host star is aligned with the core of the optical fiber once the vortex has been inserted in the pupil. Note that this mode does not provide information on the location of the companion around the host, though some insight can be extracted from the companion's measured radial velocity. 

A charge 2 K band vortex mask was installed on KPIC in 2022 as part of the phase II upgrades. The KPIC charge 2 VFN mode and its commissioning was described in great detail in a previous paper~\cite{Echeverri_VFN2023}, so we refer readers to that work for additional information on how this mode was implemented, how it works, and its on-sky performance. A separate paper also reported the first on-sky detection's from this mode~\cite{Echeverri_VFN2024}. With the charge 2 KPIC VFN mode established, we added a charge 1 vortex mask in 2024, also operating in K~band. This doubles the peak off-axis throughput for VFN at even smaller separations~\cite{Echeverri_VFN2023}. After several delays in manufacturing, we received our custom charge 1 vector vortex mask in mid February 2024 (BEAM Co.), just one month before the service mission. We quickly tested this new vortex in the laboratory at Caltech to validate key metrics and found mixed results. 

First, the transmitted wavefront error through the full optic is around 100~nm RMS, with the bulk of the error in defocus. Neglecting defocus, the optic has around 32~nm RMS. This level of wavefront error, even with the defocus, can be readily managed with the static aberration correction from the BMC DM, which has a peak-to-valley stroke of a few microns. Additionally, the optic only has three noticeable defects within the aperture, which have negligible impact for VFN given that the mask sits in the pupil and is therefore less sensitive to localized defects. 

However, the vortex mask missed the specification on multiple other properties. The monochromatic throughput at $2.0~\mu$m was around 96\% as measured with a laser. However, over a 50~nm bandwidth centered at $2.3~\mu$m, the throughput was 68\%. This is far below our specification of ${>}91$\%. The vendor carried out additional tests on a different sample and confirmed that the glue used to construct the mask has deep absorption features starting around $2.3~\mu$m, and that the absoprtion is consistent with the net loss in our mask. Furthermore, we placed careful specifications on the parallelism and wedge angle of the two glass substrates used in the vortex mask, to account for co-propagating ghosts and hence mitigate fringing while also minimizing dispersion that would adversely affect the null. One substrate was meant to be extremely parallel (${<}3''$) while the other was to be wedged at $75\pm5''$. %The goal was to achieve an effective wedge angle of ${\sim}75''$ so that the VFN ghost would land ${\sim}4.8~\lambda/D$ off-axis. 
The glass substrates met these specifications, but it seems that in gluing the two together with the liquid polymer layers for the vortex in between, the vendor ended up with one substrate tilted with respect to the other such that the effective wedge angle through the full optic is around 38$'$ (rather than the goal of 75$''$). Such a large wedge results in significant dispersion within the K~band that will limit the mask's ability to reach deep polychromatic charge 1 nulls on-sky. For additional information on the manufacturing specifications, and their implications when missed, refer to Sec.~8.2 of Echeverri~2024~\cite{Echeverri2024_PhDThesis}.

Given the late delivery, we were unable to get a new mask manufactured or adjust the existing one. However, despite the missed specifications, the vortex was able to deliver some promising nulling results in the laboratory. The null depth (i.e. stellar rejection) was about $1.5{\times}10^{-3}$ using the light source with 50~nm bandwidth at $2.3~\mu$m. This null is worse than we would have liked, especially over such a small bandwidth, but it is still sufficient for some science cases. Additionally, we achieved a laboratory peak off-axis coupling of 17.2\% simultaneous to that on-axis null. This off-axis coupling is very close to the expected ${\sim}20$\% from a charge 1 VFN system, which is promising. 

To summarize, the mask disappointingly missed specification on several key metrics. The most limiting metric will likely be the low 68\% throughput at longer wavelengths, since the CO~bandhead starts at $2.3~\mu$m and is a key feature in K~band that KPIC regularly targets for science. The next biggest limitation will likely be the dispersion from the incorrect effective wedge, which will reduce how broadband of a null we can achieve. However, the mask still provides some benefits over the charge 2 mask in specific science cases, as shown in Sec.~\ref{sec:VFNDemo}. Given that both masks are now available on the instrument, we can choose which mask to use on a case-by-case basis. 
    
\subsection{Atmospheric Dispersion Corrector} \label{sec:ADCDesign}
\begin{figure}[htbp]
\centering\includegraphics[width = \linewidth]{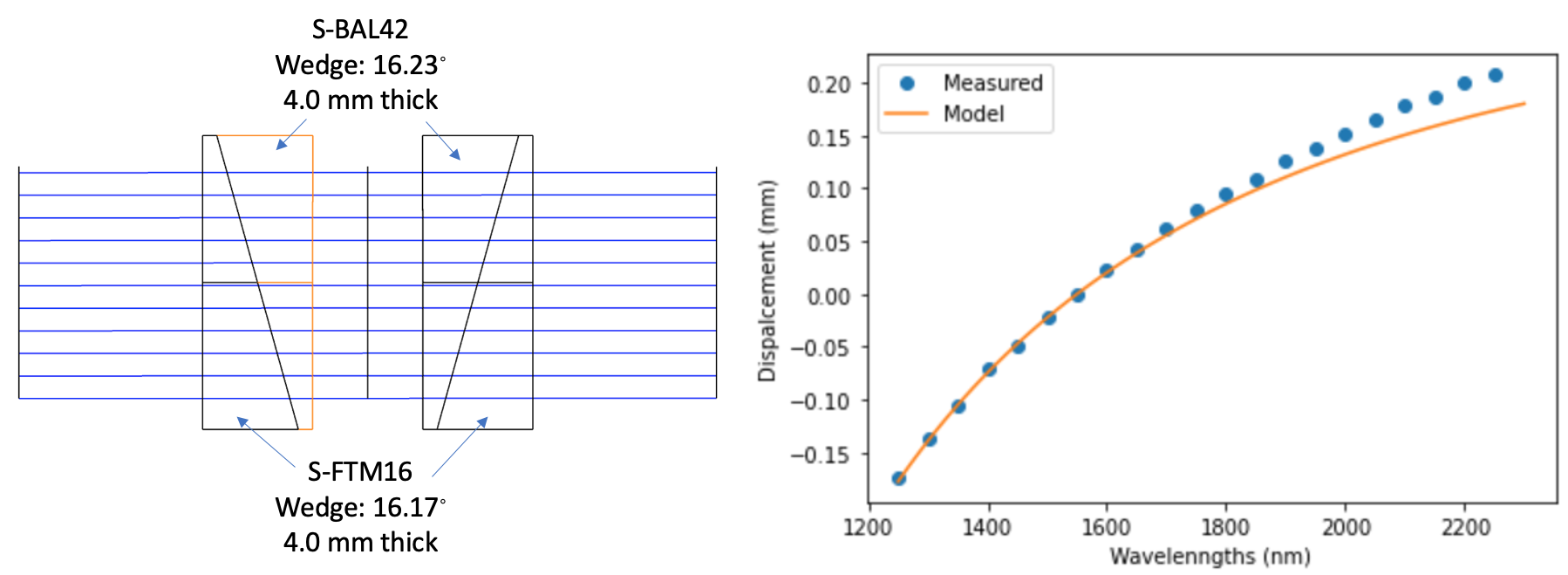}
\caption{(Left) A schematic of the two glass, two prism ADC showing the orientation of the prisms and wedge angles. (Right) The measured and modeled dispersion of the prisms. At 2200 nm, 1 $\lambda/D$ of the PSF in the experiment was 0.17 mm.}  \label{fig:ADC}
\end{figure}

The atmospheric dispersion corrector consists of two counter-rotating prisms. Each prism comprises of two wedged pieces of glass that are glued together (Fig. \ref{fig:ADC}). The atmospheric dispersion corrector is used to reverse the dispersive nature of the atmosphere and restore the beam in the focal plane to a compact point spread function. The strength of the correction is controlled by the relative rotation (clocking angle) of the two prisms and the wedge angles for each piece of glass.  This helps maximize broadband coupling into the fiber when observing a target away from zenith and is important for observing at $Y$ and $J$ band since atmospheric refraction is stronger at shorter wavelengths in the near infrared \cite{Wang2020_ADC}. The atmospheric dispersion corrector is expected to only provide a minor benefit at best at longer wavelengths because KPIC observes over a single band at a time, due to limitations in NIRSPEC. 

The main reason for the ADC is to minimize stellar leakage in the VFN observing mode. As outlined above, the ADC can be removed from the beam when its not being used. 
The VFN mode operates across K band only, but the tracking is still conducted in H band (J could also be used if desired). As such the ADC optics were optimized for performance across J to K band. The design methodology is described in\cite{wang_AAD_2020}, which refers to an early version of the KPIC ADC  that used three different glasses instead of two. After following that methodology, we optimized the residual DAR across J-K band at elevation angles between 0 and 60 degrees, by allowing two materials per prism, and varying the wedge angles of each piece of glass and clocking angle about the dispersive (elevation) axis. The angle of incidence of the input and output beams to the first and second prism pairs were also constrained to normal incidence for convenience and only materials with high transmission across the entire wavelength range were allowed. The optimal solution was obtained with a prism pair that consisted of an S-FTM16 prism with a wedge angle of $16.17^{\circ}$ and an S-BAL42 prism with a wedge angle of $16.23^{\circ}$. The configuration of the prisms is shown in the left panel of Fig.~\ref{fig:ADC}. With this combination the residual DAR could be theoretically reduced to 2 mas across $K$ band ($\sim1/20$th of the width of the point spread function).  

The prisms were fabricated by Opticraft. They were characterized in the laboratory when they arrived and the throughput of the entire ADC (2 prism pairs) was confirmed to be $\sim95\%$ across J-K band. The transmitted wavefront error through the ADC with power removed was $<27$~nm. The dispersion was measured and is shown compared to the theoretical dispersion curve in the right panel of Fig.~\ref{fig:ADC}. It can be seen that there was a departure from what was expected. For reference, 1 $\lambda/D$ at 2200 nm is 0.17 mm in the experiment. Therefore, the error of 0.05 mm at the longest wavelengths would correspond to 0.3~$\lambda/D$ or 12.3 mas.  

We believe that this just means we will have to use different clocking angles, and may get slightly worse performance than anticipated, but needs to be confirmed through simulation. We also discovered that the prisms steer the beam orthogonal to the dispersion axis as well. This implies that the optics have an effective wedge in that direction which is non-negligible. After some discussions with the vendor, it was determined that there was a clocking error between the two prisms at the time of bonding. This added some wedge to the axis that was not supposed to have any and modified the wedge for the dispersive axis, which probably explains why the dispersion does not match theory. With limited time before deployment, we decided to use the prisms in hand. 

Two prisms were selected with the minimal beam walk orthogonal to the dispersive axis. From measurements in the laboratory, both prisms will walk the beam with a radius of $<$12~$\lambda/D$ on the tracking camera in H band, as the prisms are rotated. The tracking loop will be used to compensate for this beam walk and keep the beam pointing on the optical fiber. The tracking loop has sufficient bandwidth to correct for this because the ADC prisms rotate very slowly.

\subsection{PIAA} \label{sec:PIAADesign}
The PIAA optics are used to boost coupling to the optical fiber by apodizing the flat topped pupil illumination into a soft edged beam which is a better match to the fiber mode. The design and performance of the KPIC PIAA lenses is summarized in~\citenum{calvin_EDE2021}. Although optimized across K and L band, the biggest coupling efficiency gains come in L-band (boost from 65\% without PIAA to 86\% with PIAA). The PIAA uses a lossless apodization which makes it ideal for L band observations where the thermal background is higher. The PIAA does not improve speckle suppression. To use the PIAA, it is translated into the beam where it has been pre-aligned with one optical fiber. Then once the target is acquired, it is steered to that fiber.

\subsection{Zernike Wavefront Sensing}
Although not core to KPIC operations, a Zernike WFS was installed in March 2022. A vector Zernike mask based on a metasurface optic was installed near the focal plane of the first lens in the tracking camera in the pupil viewing optics arm. In this way, when pupil viewing mode is selected, the TTM can be used to steer the beam precisely onto the phase dimple and convert the pupil viewing mode into a WFS. To take full benefit from the vector nature of the mask, a Wollaston prism was installed in the pupil viewing relay to generate two images, one for each linear polarization state. The Zernike WFS has been demonstrated to be capable of correcting for static aberrations with high fidelity off-sky as well as for compensating for segment piston errors in the primary mirror of Keck~\cite{Salama_KPM2024}. For more details, please see~\citenum{Salama_KPM2024}. The ZWFS can not be used in routine KPIC operations as the tracking camera is needed to track on the target.

%\subsection{New Observing Modes}
%\label{sec:new_modes}
%With the upgrades and modes listed above, KPIC has expanded observing capabilities as described below. 
%\begin{itemize}
%    \item \textbf{Direct spectroscopy -} 
    %\item \textbf{Apodized direct spectroscopy -} the microdot apodizer is designed to eliminate the majority of the static speckles from diffraction in the focal plane from 3-12 $\lambda/D$, thereby improving the detection of companions at these smaller separations. It does this at the expense of throughput, and with the increased losses, emissivity becomes an issue. As such the microdot apodizer was optimized for observations at shorter wavelengths (K band). To use the microdot apodizer, the mask is first inserted into the beam, and then the TTM steers the known companion onto the fiber of choice. 
    
%\end{itemize}

%%%%%%%%%%%%%%%%%%%%%%%%%%%%%%%%%
\section{Laboratory Validation}
\label{sec:lab_validation}
Before deploying to Keck, we tested the phase II system in the laboratory against a comprehensive list of success and pre-ship criteria and milestones developed to assess the performance of each sub-module, as well as the performance of the instrument as a whole. In this section, we report on some of the throughput and coupling efficiency measurements of the system. Note that this is just a small subset of the key results obtained in the laboratory. 

The throughput at 2~$\mu$m (a proxy for K band) from the input fiber focal plane to the fiber bundle focal plane was measured to be 72.6$\%$ without the PIAA and 69$\%$ with the PIAA. Note that this is the transmission of the optics up until the fiber bundle, not the coupling efficiency into the fiber; the losses with the PIAA in the beam path are due to the added optical surfaces but they are balanced out by the increased coupling efficiency from the mode matching that the PIAA provides. We can derive a theoretical throughput value from a bottom-up analysis where we account for the theoretical throughput of each optic. This yields an expected throughput of 5$\%$ higher than what was measured. Therefore our measured value is highly consistent with expectation. In addition, we measured the throughput from the input focal plane to the tracking camera with a 1550 nm laser to be 14$\%$ with the KL long pass dichroic in both the PyWFS and tracking camera pickoff mechanisms, also consistent with expectation to within error. The choice of test wavelengths is consistent with the baseline direct spectroscopy observing mode where we observe in K and L and track in H band. Given most optics in the FIU are reflective gold, and the yH lens throughput was $\sim$99\%, we can expect similarly high levels of transmission for the shorter wavelengths taking into account the relative throughput's for various dichroic profiles shown in Fig.~\ref{fig:dichroicsfilters}. 

\begin{figure}[htbp]
\centering\includegraphics[width = 0.8\linewidth]{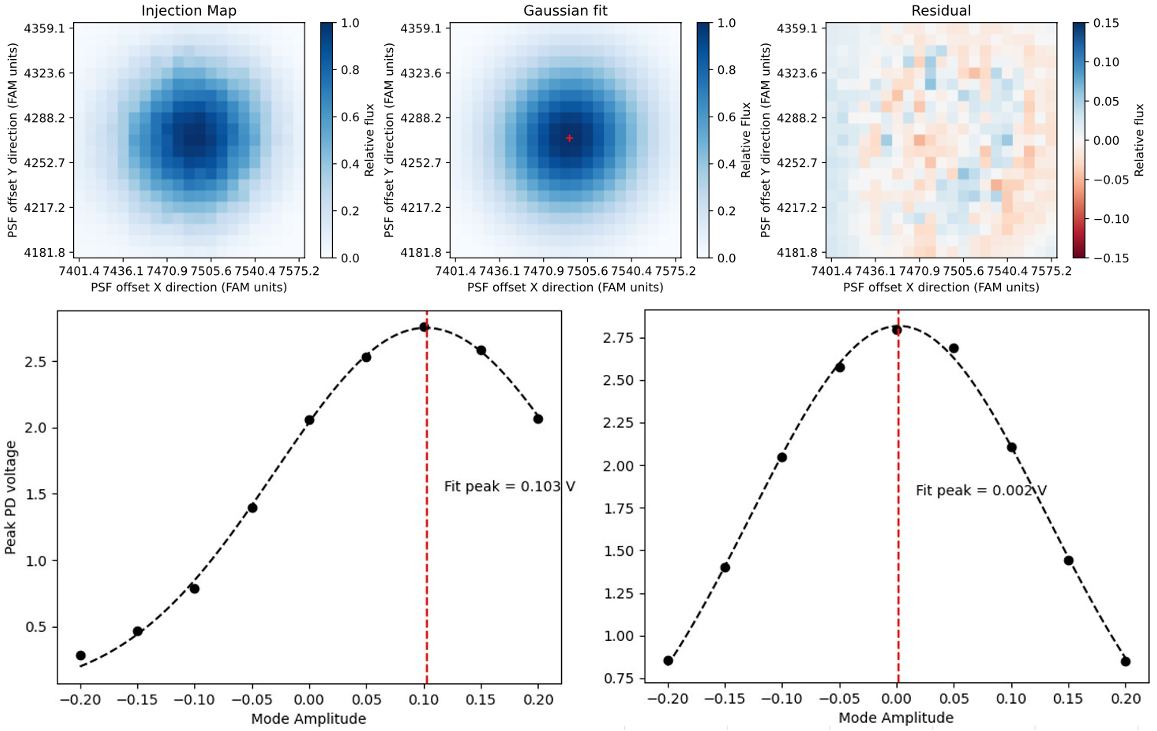}
\caption{(Top) Coupling maps made by scanning the TTM and recording the flux on the PD. (Left) Measured, (Center) Gaussian fit, (Right) Residual difference between measurement and fit. The residuals are only used as a visual check of the goodness of fit. (Bottom) Flux measured on the PD as the defocus Zernike was scanned. (Left) Before correction, showing that some defous was present at the time. (Right) After applying the defocus correction there is a negligible amount of defocus left.    \label{fig:maps}}
\end{figure}

We then measured the coupling efficiency into the KL bundle. Initially we placed a flat mirror in the location of the DM for this experiment, which allowed us to measure the coupling efficiency as limited by internal static aberrations. A PD was placed at the output of the bundle and the signal was sampled by the KPIC control computer. We conducted a Tip/Tilt (TT) scan with the mirror across the core of the fiber and measured the transmitted power. We fit this coupling map with a 2D Gaussian to find the peak and its location as seen in the top row of Fig.~\ref{fig:maps}. 
As the multiport provides the ability to adjust the distance between the focusing lens and the bundle, we made small adjustments to this gap manually and repeated the TT scan until we found the plane of best focus. The optimal coupling efficiency at 2~$\mu$m was measured to be 71-73$\%$. Given the pupil is unobstructed when using the internal source of KPIC, the theoretical limit is expected to be $\sim$80$\%$\cite{Shaklan1988}. After the flat mirror was replaced with the DM, we scanned the amplitude of the first 10 Zernike modes while monitoring the power on the PD. For each amplitude we applied for each Zernike, we conducted a full TT scan to compensate for any beam walk that may have occurred by applying the aberration. For each mode we plotted the peak flux from each coupling map as a function of the Zernike amplitude applied, and fit the the data with a Gaussian to find the optimum aberration amplitude to maximize coupling (see bottom row in Fig.~\ref{fig:maps}). We stepped through each Zernike mode in turn and improved the coupling efficiency up to 75-77$\%$. This demonstrated that we could achieve near perfect coupling efficiency in the laboratory with very little static wavefront correction required from the DM, validating that the optical beam train was aligned with very low wavefront error. 

Next we validated the stability of the injection, which is key to making high quality observations for a full night. 
\begin{figure}[htbp]
\centering\includegraphics[width = 0.8\linewidth]{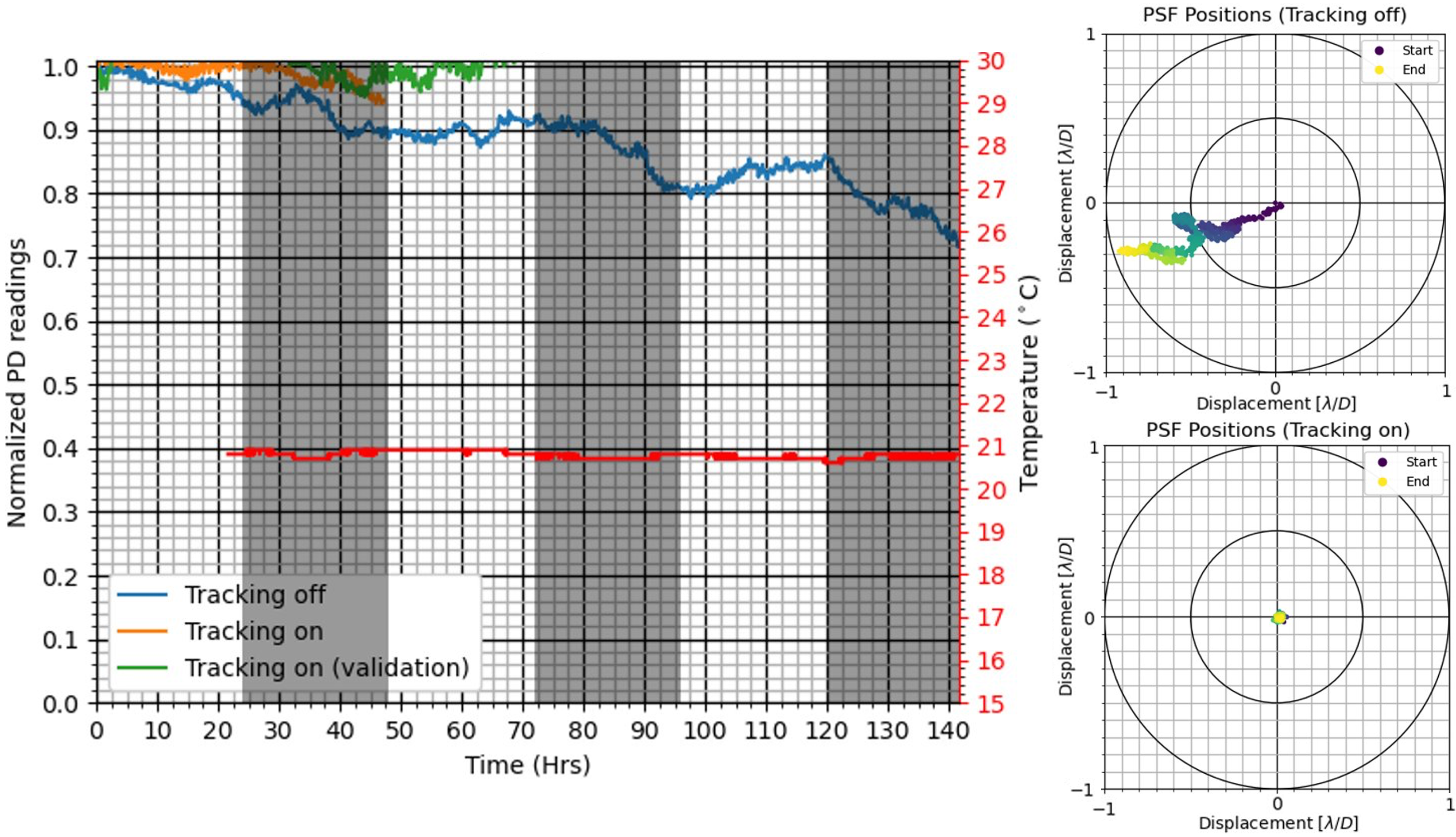}
\caption{System stability over the course of several days with and without tracking control loop active. (Left) Normalized PD readings through the bundle. Blue curve is with tracking control disabled while green and orange are repeat samples with tracking enabled. White and gray shading delimits 24-hour periods. (Right) PSF location on tracking camera relative to the start position with tracking loop off (Top) and with tracking loop on (Bottom).   \label{fig:stability}}
\end{figure}
With the coupling to the fiber optimized using the techniques described above, we tracked the flux evolution on the PD over a period of 6 days. The blue curve in the left panel of Fig.~\ref{fig:stability} shows the result. Over the course of the 6 days with no active control, meaning the system was allowed to drift and settle uncompensated, the power in the PD reduced by about 25$\%$. The location of the PSF is shown in the top right panel over that same period. It is clear that there is a drift of about $1\lambda/D$ in the location of the PSF at the tracking camera and this can be correlated with the drop in the flux. The temperature was measured in the vicinity of the instrument and is shown in red. The temperature varied by 0.2$^{\circ}$C peak-to-peak during the test and although difficult to see, does not correlate with the observed drift to first order. Nevertheless, given that an observing night is no longer than 12 hours, this demonstrates an extraordinary level of stability for the size of the fiber mode ($\sim$10 $\mu$m); over the first 12 hours of the test, there is a drop of less than 4$\%$. Further tests were conducted with KPIC's tracking loop which uses the tracking camera to determine the location of the beam and drives the TTM to actively maintain the alignment of the beam with a given pixel on the detector. The resulting PSF stability on the tracking camera with tracking control on is shown in the bottom right panel indicating no deviation in the beam. The flux during this test, which was repeated twice, is shown in the green and orange traces in the left panel and although it fluctuates by a few percent, maintains very high flux levels ($>$95\%) once the tracking loop was closed using the tracking camera. This indicates that there is very little non-common path drift between the tracking camera and the fiber bundle, and hence we can calibrate the two before a science run and drive the system to maximal coupling for an extended period thereafter.

%%%%%%%%%%%%%%%%%%%%%%%%%%%%%%%%%
\section{Enhanced Calibration Procedures}
\label{sec:improved_cal}
In phase I, both fiber finding and static aberration corrections relied on a source being scanned across the fiber tip (TT scan), while flux was being recorded on the slit viewing camera of NIRSPEC. The slit viewing camera of NIRSPEC has very bad cosmetics which vary with time making it challenging to extract flux accurately and reliably. In addition, we had issues with the open loop calibration of the Keck AO Xinetics DM. In phase II we therefore replaced the slit viewing camera with a PD installed in the FEU (as outlined above) and utilized the 1k BMC instead. These replacements, along with optimization of the laser signal from the 2~$\mu$m laser have improved the quality and sped up the daytime calibrations procedures. 

\begin{figure}[htbp]
\centering\includegraphics[width = \linewidth]{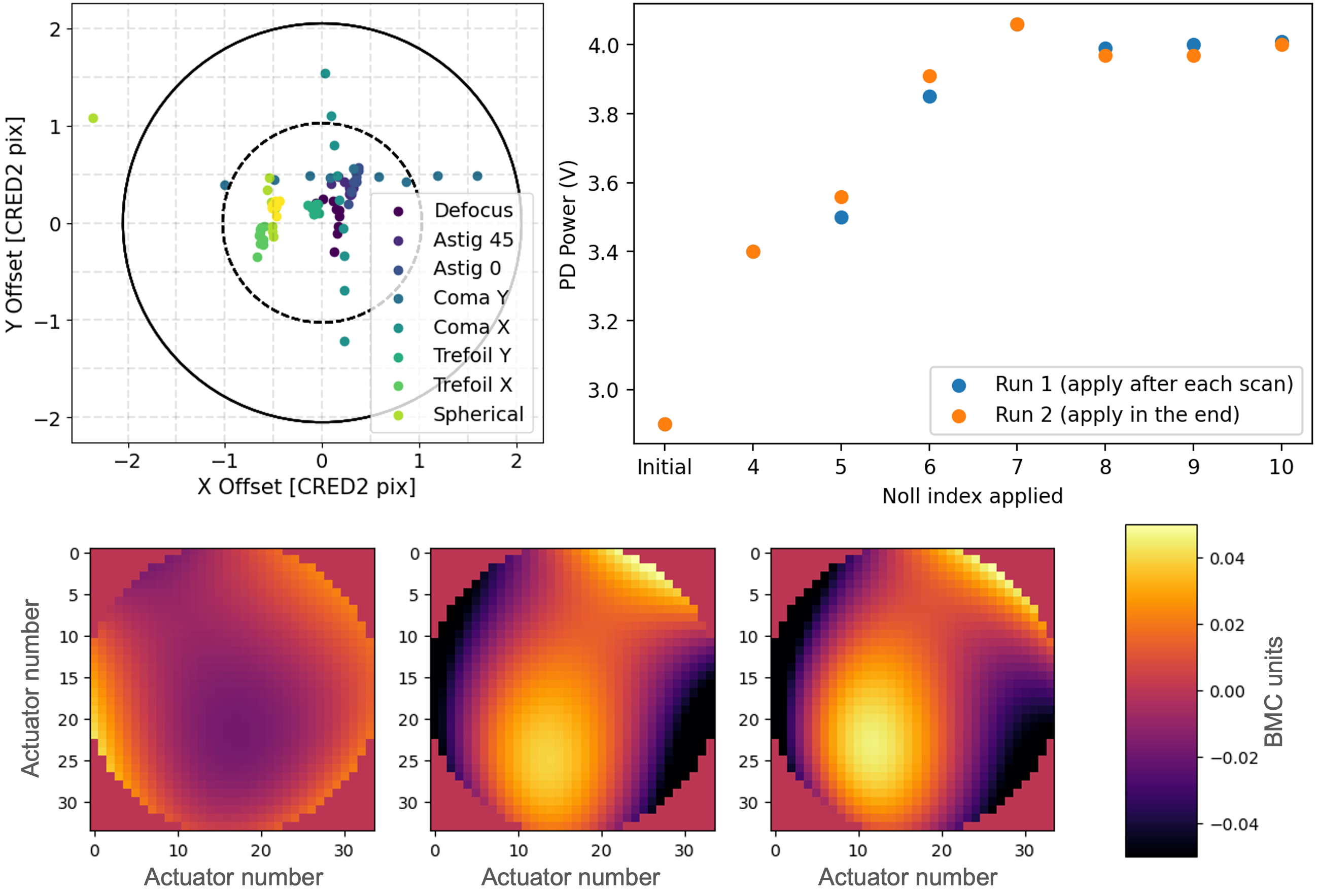}
\caption{Results from Zernike scan tests in the laboratory. (Top left) The centroid for the PSF on the tracking camera as Zernike amplitude was scanned. (Top right) The relative improvement in the flux on the PD when scanning Zernike modes applying the recommended Zernike amplitude between each successive mode and after the entire scan has been completed. (Bottom) DM shapes before starting a Zernike scan (left), after completing a scan and applying each Zernike amplitude in turn (middle) and applying all the recommended Zernike amplitudes at the conclusion of the experiment (right). The DM maps and measured flux levels are very similar.   \label{fig:NCPAs}}
\end{figure}

With this new setup, we investigated whether the Zernike modes were truly orthogonal in our system. The motivation for doing this was to reduce calibration time by for example being able to remove the TT scan for each Zernike amplitude. To do this, we conducted two tests. The first test involved scanning the amplitude for each Zernike term from -0.2 to +0.2 BMC units (1 BMC unit is equivalent to the full deflection of the DM, which is $\sim$3.7~$\mu$m of stroke) up to the spherical aberration term, and computing a TT scan while measuring the flux on the PD at each setting. The flux data were fitted with a 2D Gaussian and the center extracted. The pixel on the tracking camera at which the peak in coupling occurred is shown in the top left panel of Fig.~\ref{fig:NCPAs}. It can be seen that the centers of the coupling maps remain localized around the 0, 0 coordinate on the tracking camera except for the two coma aberrations which display linear trends with varying amplitude and spherical aberration which is clustered but offset. The linear trend as a function of the amplitude of the coma aberrations are to be expected given the morphology of the PSF when coma is present. The PSF appears to shift laterally which increases with increasing amplitude. The offset nature of the spherical aberration points is unknown. To be able to interpret this data, its essential to realize that the plate scale of the tracking camera is 7.183 mas/pixel. Given a PSF with a FWHM of $\sim$40 mas at 2~$\mu$m, the majority of the centroids for small amplitude Zernike aberrations are localized within $\pm$0.5 pixels or $\sim$3.5 mas ($\sim$1/10$^{th}$ of the PSF) of the starting coordinate. From this result we concluded that if the starting Strehl for the beam delivered by the AO system to KPIC is low (which was the case when we first installed KPIC inside the AO bench in 2022), then TT scans should be conducted on at least the coma modes during the Zernike optimization process to maximize the coupling improvement. However, once a DM map is in hand that corrects for the bulk of the static aberrations such that the starting Strehl is $>50\%$, the small amplitude aberrations present in the AO+KPIC beam trains from drifts and temperature fluctuations on a daily basis can be well compensated without the need for TT scanning at each Zernike mode amplitude. Given that each TT scan consisted of 7$\times$7 data points, this resulted in an improved calibration script which was 49$\times$ faster.  

The second experiment we conducted to confirm orthogonality and determine how many modes we should correct relied on running the Zernike scan sequence (TT scan free) and comparing the measured flux and final DM map solutions when the optimum Zernike amplitudes for each term were applied after each successive mode and/or at the end of all the scans. The results are shown in the top right and bottom panels of Fig.~\ref{fig:NCPAs}. It can be seen that regardless whether the optimum Zernike amplitude is applied immediately following the scanning of a given mode or recorded and applied after all modes have been scanned, the improvement in flux evolves in a similar manner. This can only be true if the modes are orthogonal. The final DM map solutions to compensate for the static aberrations are nearly identical for the pre and post application of correction cases. Note, the scale bar ranges from $\pm$0.05 BMC units, validating less stroke is needed to compensate for AO bench+KPIC aberrations after an initial map has been determined. An interesting outcome of these tests is that we determined that there was no benefit of scanning Zernike terms beyond a noll index of 11. This allowed us to eliminate higher order modes from our scans, again improving static aberration compensation times. The final scanning process now includes scanning 8 Zernike modes (Nolls 4 through 11), with 21 amplitudes from -0.3 to +0.3 BMC units (with 0.03 BMC unit steps).

With the improvements outlined above, the Zernike scanning process is now routinely completed in $\sim$5 minutes. This improvement in efficiency means we can now make optimal DM maps for each observing mode efficiently and load them when we switch between modes during the night. 

% It should be pointed out that utilizing the flux through the fiber to drive the DM was the core building block for the demonstration of speckle control in phase II, albeit using the NIRSPEC detector for sensitivity reasons~\cite{Xin_OSS_2023}.

We have tracked the evolution in the static aberration compensation solutions over the period of an observing run ($\sim$1 week) several times. We find small amounts of drift from day-to-day. This builds on the laboratory measurements in Fig.~\ref{fig:stability} which validated small drifts internal to KPIC by making a similar conclusion with all optics in the AO beam train. The penalty in coupling for not re-scanning Zernikes each day is minimal, but given the efficiency of the scan, it is typically repeated on each day of observing to maximize performance.  

The same TT scan used for optimizing alignment of the PSF onto the SMF is used to find the minimum in the null in the VFN mode, just with a finer scan composed of more and smaller steps. The procedure that minimizes aberrations however is more complicated since it must compensate for aberrations in the vortex mask and since the VFN mode is more sensitive to specific aberrations. As such, we do additional NCPA scans focused on improving the null depth by re-scanning for coma for charge 1 and astigmatism for charge 2 with the corresponding vortex mask in the beam. Finally, we do an additional set of coma and astigmatism scans but using NIRSPEC as the detector to optimize the null specifically at 2.3~$\mu$m (the CO bandhead), since the PD scans are done with a 2.0~$\mu$m laser which is far from the science wavelengths.

%%%%%%%%%%%%%%%%%%%%%%%%%%%%%%%%%
\section{Commissioning Results}
\label{sec:commissioning}

\subsection{Calibrating Plate scale, 
 and Orientation}
Following the upgrades initiated in 2022 and finalized in 2024, the base operations of KPIC were verified and restored so science could recommence. Firstly, the plate scale, distortion solution and orientation of the tracking camera with respect to North had to be updated. The process outlined in \citenum{Delorme2021_KPIC} was used. This involved imaging a range of low contrast binaries on NIRC2, which has an extremely well known plate scale and distortion solution to determine the separation and orientation, before taking a series of images of the binary system at various locations on the tracking camera. To move the targets across the tracking camera, the internal TTM in KPIC was used. Images of the binaries at different locations on the detector allow the distortion solution to be determined across the entire detector accurately. This method has been repeated numerous times over the 2 years to track its stability and to see how small servicing of the instrument impacts it. The plate scale as of April 2024 is 7.18 mas/pixel$\pm$0.3\%. The North angle is 85.9$\pm$0.2 degrees. The uncertainty in the offset would be $\pm$3 mas at 1 arcsec separation due to plate scale which corresponds to 0.16 $\lambda/D$ at 1 $\mu$m and 0.08 $\lambda/D$ at 2 $\mu$m. For a rotation error of 0.2 degrees at 1 arcsec, the offset would be $\pm$3.5 mas, which would result in an error of 0.19 $\lambda/D$ at 1 $\mu$m and 0.09 $\lambda/D$ at 2 $\mu$m. At 1 $\mu$m, the loss associated with such errors would be 10\% loss. At 2 $\mu$m the loss due to these offsets would be $<$5\%. As a point of reference the plate scale was 7.24 mas/pixel and 85.9 degrees back in August 2022. These are small changes, but when observing targets at 2 arcsec can cause misalignment's with the optical fiber and a reduction in efficiency so we track and update them every few months. 

\subsection{Validating Tracking Loop Functionality}
\begin{figure}[htbp]
\centering\includegraphics[width = \linewidth]{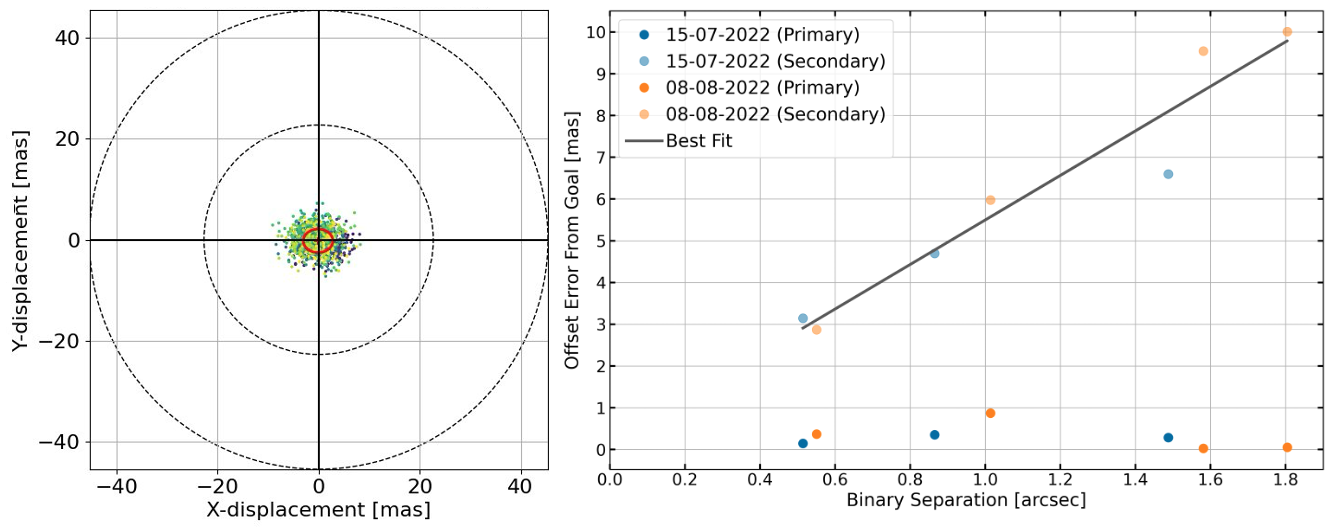}
\caption{(Left) XY scatter plot of centroid locations from 1000 frames acquired with the tracking camera with the tracking loop closed. The distribution is centered on the 0,0 coordinate which was the intended target validating that the tracking loop is working. (Right) Offset error in centroids for tracking on the primary and secondary of several systems as a function of separation.     \label{fig:tracking}}
\end{figure}

We next tested that the tracking loop was functioning. This entails setting the tracking camera integration to have a PSF with a signal to noise of at least 50 and maximizing the frame rate. The tracking loop relies on several C-red2 frames being averaged, the centroid of the ensemble determined, and using this information to drive the TTM to maintain the alignment of the target with the pre-determined sub-pixel position corresponding to the fiber of choice. The frame averaging makes the tracking loop a drift compensation loop rather than a fast TT loop. Figure~\ref{fig:tracking} left shows the centroid distribution collected from 1000 consecutive images about the desired location on the C-red2. It can be seen that there is a Gaussian like distribution centered on the target pixel. This has been validated on many targets since and demonstrates the tracking loop is functioning as desired. 

\subsection{Validating Plate Scale and Distortions}
To validate the plate scale/distortion solutions, a series of low contrast binaries were observed with the C-red2. In each case the tracking loop was closed on the primary. After several cubes of 1000 frames were collected in this configuration, the off-axis companion was tracked on and another set of data was acquired. The centroids for both were analyzed to see how close to the target pixel the primary and secondary landed. Figure~\ref{fig:tracking} right shows the offset distance (computed as the square root of the sum of the squares of the offsets in x and y) as a function of the separation of the companions observed. Unsurprisingly, the primary of each system has very little offset from the target pixel location which is consistent with the tracking loop functioning as expected. However, the companions have some residual offset, which increases linearly with separation. From this data set collected after the initial install of KPIC phase II in 2022, the residual offset direction was determined to be in the direction of separation. This hints at a small residual error in the plate scale which reaches up to 25\% of the PSF FWHM in K band for targets separated by 1.8~arcsec. However, the interesting targets are within 0.2-arcsec where the offset error would be$<1$~mas, which has a negligible impact on coupling at all wavebands. After the 2024 service mission, the experiment was carried out once again and a similar trend and magnitude of error were confirmed, but this time it appears the residual error is mostly orthogonal to the offsetting direction. Although this change of direction in residuals is unknown, we note that the level of coupling loss due to misalignment's of this magnitude at small and large separations is tolerable.

Given the ADC is not baselined for all KPIC observations, it is critical to compensate for the offset between the PSF locations in H band, which the tracking camera sees and K band, where the injection unit operates in some modes when operating away from Zenith. This is based on a model of the differential atmospheric refraction (DAR) which is a function of the Zenith angle (amongst other parameters)~\cite{Mathar_2007}. The TTM offsets the PSF on the tracking camera along the elevation axis of the telescope, so that the K band PSF is aligned with the fiber. By adding manual offsets in the elevation axis while recording data on NIRSPEC, we plotted the flux on NIRSPEC vs elevation axis offset. We confirmed that the maximum flux does indeed occur at the offset predicted by our model validating the DAR compensation is working. 

\subsection{End-to-end Throughput \& Science Validation}
\begin{figure}[htbp]
\centering\includegraphics[width = 0.8\linewidth]{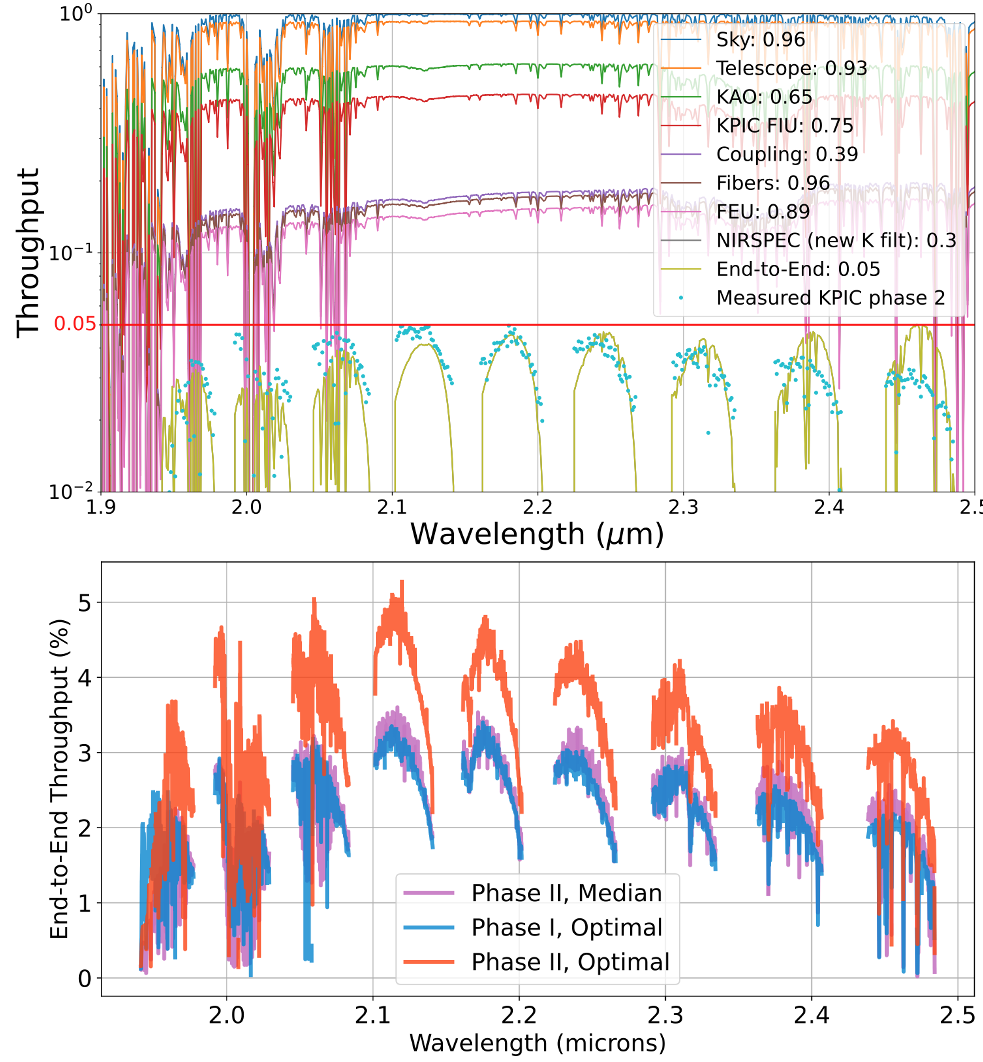}
\caption{(Top) The throughput budget for the KPIC science path including the measured throughput on-sky for target KELT9 on 2022-07-23. The estimate of the terms in the throughput budget come from a combination of theoretical value, measurements and extrapolations from measurements. (Bottom) The optimal throughput for the phase I instrument (Kappa And, 2020-07-03) compared to the median and optimal (KELT 9, 2022-07-23) performance of the phase II instrument. The optimal phase I performance is equivalent to the median phase II performance. Note, the median performance curve in the was acquired by observing dozens of targets over a range of conditions over 7 nights in July and August in 2022.   \label{fig:throughput}}
\end{figure}

We next measured the throughput from sky-to-detector. The end-to-end throughput is computed as the ratio between the number of photons measured on the detector, and the number of photons expected for the star, given its K band magnitude. The former is computed by extracting the flux in data numbers across each spectral order and multiplying the detector gain (3.03e$^{-}$ / data numbers). The latter is computed using a template spectrum of the star, flux normalized to its K band magnitude. We factor the integration time, telescope size, and wavelength range of each spectral bin for the expected number of photons. Here, the size of each spectral bin is set by the width of the line spread function at that wavelength and is roughly 2.5 pixels on average. Figure~\ref{fig:throughput} top shows the measured throughput on KELT9 taken on 2022-07-23, overlaid with the throughput contributions from various aspects in the optical path. The bottom panel shows that the throughput reaches $\sim$3\% around 2.1$~\mu$m in median conditions, and close to $\sim$5\% in the best conditions (same data used in the top panel). This is a significant improvement from the best that was seen in phase I, which was $\sim$3\% only in optimal seeing conditions (acquired on Kappa And on 2020-07-03). Indeed, the performance of the phase II instrument in median conditions matches the performance of the phase I instrument in optimal conditions. Note, the median performance curve in the bottom panel was acquired by observing dozens of targets over a range of conditions over 7 nights in July and August in 2022. We attribute this to the superior static aberration correction due to the enhanced daytime calibration procedures and the new NIRSPEC filter. The largest contributor to losses is unsurprisingly the coupling term. This is expected in an SMF injection unit as the coupling scales linearly with Strehl. Given the peak coupling efficiency for the Keck pupil geometry to an SMF is 65\% (aberration free), then a 55\% K band Strehl ratio would be needed to achieve the throughput's observed, which is consistent with Keck AO performance in the conditions on the night. 

\begin{figure}[htbp]
\centering\includegraphics[width = \linewidth]{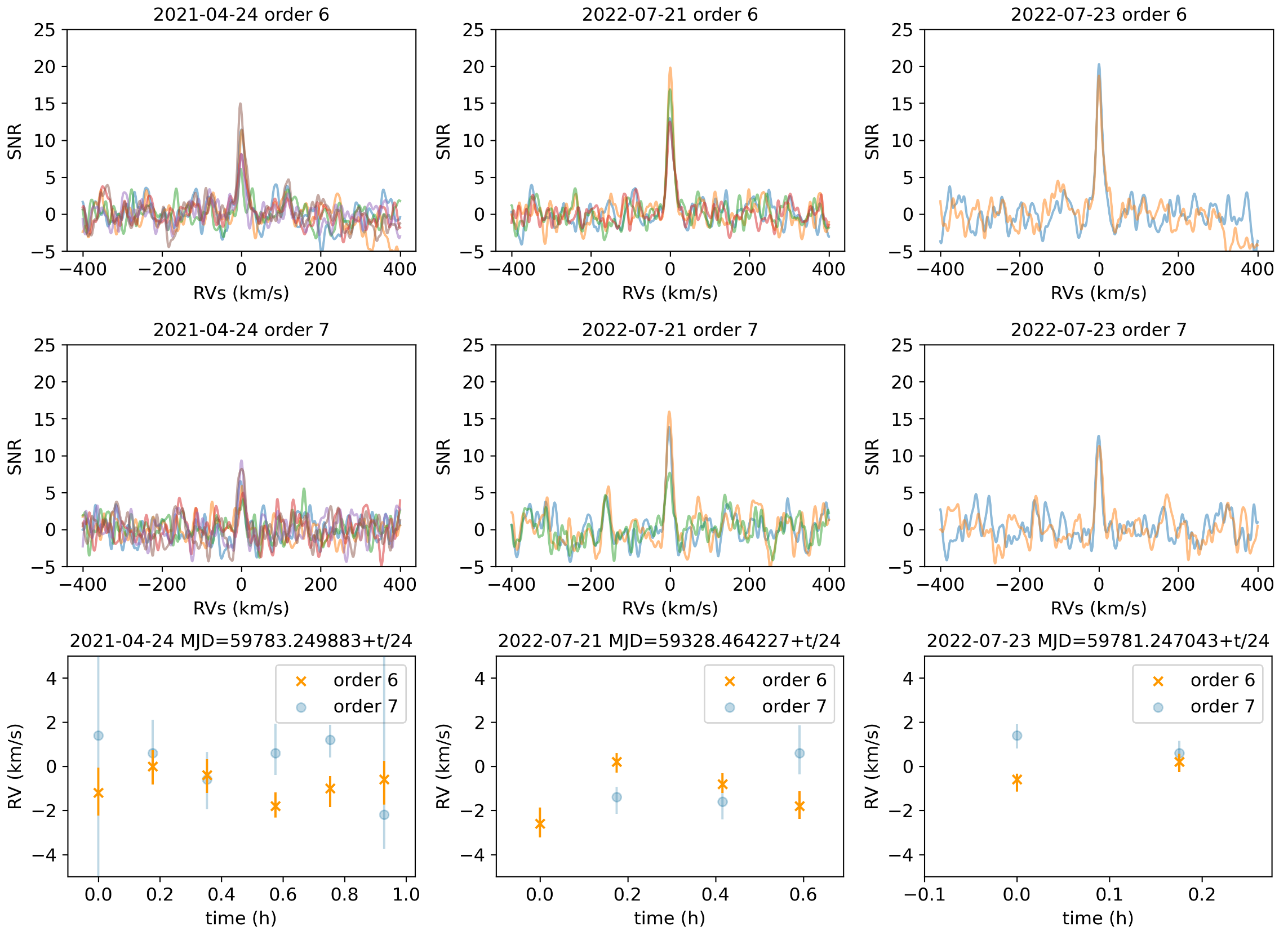}
\caption{(Top) and (Middle) show cross-correlation functions between the measured spectrum of GQ Lup B and a template spectrum. The peaks identify a strong detection. The height of the peak indicates the SNR of the CCF detection. Each trace corresponds to an individual 10 minute exposure taken with NIRSPEC. Orders 6 and 7 of NIRSPEC correspond to the region around the CO band head. The SNR for the observing dates in 2022 with phase II is superior to the 2021 phase I result. (Lower) Shows the measured radial velocity from each frame with error bars. The error bars are much reduced with the phase II variant of the instrument.    \label{fig:CCFs}}
\end{figure}

Finally, we collected data on a companion that was studied in phase I to confirm that KPIC was once again ready for science. The target we observed was GQ Lup B. Residing at an angular separation of 0.7 arcsec, GQ Lup B is sufficiently far from the star that the limiting noise source is thermal background noise from the spectrograph~\cite{Wang2024_KPICContrastCurve}. Thus, an increase in throughput on the companion translates to a linear increase in its signal-to-noise ratio. Three observations were carried out. The first was on the 2021-04-24 using the phase I instrument. The next two observations were carried out on 2022-07-21 and 2022-07-23 using the phase II instrument. It should be noted that GQ Lup B is low in the sky, which means the high airmass results in poorer AO performance and a lower throughput compared to typical Northern hemisphere targets. The results are shown in Fig.~\ref{fig:CCFs} and summarized in Table~\ref{tab:GQLupBperformance}. The throughput improved by 50-90\%, some of which is attributed to the phase II upgrades consistent with the results above and some of it due to the better seeing on the two nights in phase II. Despite the better seeing, shorter integration times were used in the phase II observations. 
The raw data was reduced with the KPIC data reduction pipeline (DRP)\footnote{\url{https://github.com/kpicteam/kpic_pipeline}} to produce calibrated 1D spectra~\cite{Delorme2021_KPIC, Wang2021_KPICScience}. The companion is then detected by fitting a joint forward model including the companion signal, the tellurics, and a model of the starlight following the approach described in~\citenum{Ruffio-DER2023} using the Python package BREADS~\cite{Agrawal2023AJ....166...15A}. The companion cross correlation function is obtained by doppler shifting a BTSettl atmospheric model ($\log g=4.0$; $T_{\mathrm{eff}}=2700\,\mathrm{K}$, \citenum{Allard2012}) for the companion.
% The KPIC data reduction strategy relies on computing the cross-correlation function (CCF) between the reduced data and a template spectrum as outlined in~\cite{Delorme2021_KPIC}. 
The top and middle panels in Fig.~\ref{fig:CCFs} show distinct peaks in the CCFs indicating that the companion was detected. The colored traces are individual CCFs for each 10 minute exposure acquired. It can be seen that the CCFs in phase II have much higher SNRs as compared to phase I. Indeed the CCF SNR gains were 1.6 and 1.95 respectively. The radial velocity and associated errors were computed and are shown in the bottom panels of Fig.~\ref{fig:CCFs}. It can be seen that the error bars greatly reduced in phase II. Table~\ref{tab:GQLupBperformance} shows a reduction in the radial velocity error (R$_{verr}$ from 770 m/s down to 410-460 m/s. Once again, these improvements are associated with the higher throughput resulting from the upgrades and better seeing conditions. These results validate that the KPIC instrument has been recommissioned and is operational and elucidate the improved performance.   

\begin{table}
\begin{center}
\begin{tabular}{|c|c|c|c| }
%\hline
%& \multicolumn{3}{|c|}{Throughput (\%)} \\
\hline
Date & 2021-04-24 & 2022-07-21 & 2022-07-23 \\
\hline
Phase & I & II & II \\
\hline
\hline
DIMM Seeing (as) & $\sim0.9$ & $\sim0.6$ & $\sim0.5$  \\ 
MASS Seeing (as) & $\sim0.5$ & $\sim0.3$ & $\sim0.15$ \\ 
Throughput (\%) & 1.6 & 3 & 2.4 \\ 
CCF SNR & 10 & 16 & 19.5 \\ 
R$_{verr}$ (m/s) & 770 & 460 & 410 \\ 
T$_{\mathrm{int}}$ (min) & 60 & 40 & 20 \\ 
\hline
\end{tabular}
\caption{Comparison of performance of KPIC on GP Lup B in phase I and II.}
\label{tab:GQLupBperformance}
\end{center}
\end{table}

%\subsection{PIAA}
%Would be great to show its useful in L band. If we can get a data set.

\subsection{Microdot Apodizer} \label{sec:MDA}
\begin{figure}[htbp]
\centering\includegraphics[width = \linewidth]{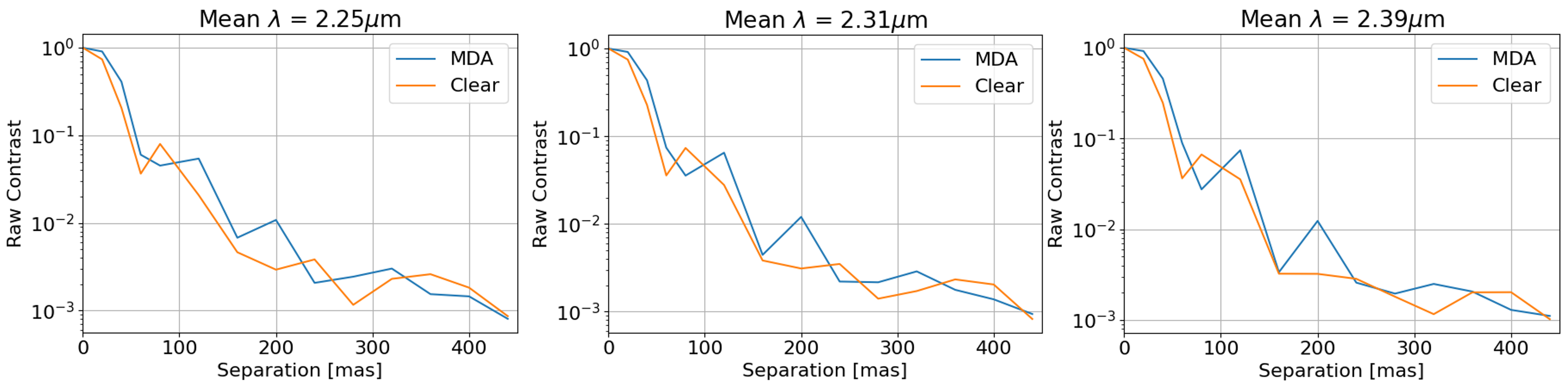}
\caption{Raw contrast with and without the microdot apodizer in the beam at 2.25~$\mu$m (left), 2.31~$\mu$m (middle) and 2.39~$\mu$m (right), on HD196885 on 2022-07-15.
\label{fig:MDA}}
\end{figure}

At deployment in 2022, a microdot apodizer was initially installed in the coronagraphic mask selector alongside the charge 2 vortex mask. For full details about the apodizer, its design, and laboratory performance, please see~\citenum{calvin_EDE2021}. The microdot apodizer was tested several times on-sky in K band in 2022. Briefly, this is a mask which comprises hundred and thousands of chrome dots on a glass substrate that apodizes the beam on the inside and outside edges of the pupil eliminating diffraction from the pupil shape. It is specifically designed to reduce diffracted light from 5-12~$\lambda/D$ in K band. Figure~\ref{fig:MDA} shows the on-sky raw contrast as a function of separation with the microdot apodizer in the beam and a clear pupil, at several wavelengths across K band. This was created by moving the fiber from the target (HD196885) to off axis locations in a linear fashion and acquiring the data with NIRSPEC along the way, on the night of 2022-07-15. There is no reduction in the stellar flux between 200 mas (5~$\lambda/D$) and 450 mas (11~$\lambda/D$), despite a factor of 10 being observed in laboratory testing. The seeing on the night was 0.91" (above median conditions) and variable during this acquisition. This experiment has been repeated several times in various conditions and the results presented are representative of all data sets. This lack of improvement in contrast combined with the low throughput of the microdot apodizer ($\sim50\%$), negates the utility of the mask. The reason for the reduced contrast improvement on-sky is believed to be due to a poorer dynamic speckle halo than originally expected, which is dominating over the quasi-statics at the spatial frequencies described above. For this reason, the microdot apodizer was removed and in 2024 at the service mission and replaced with the charge 1 vortex mask.   

\subsection{Speckle Nulling}
Speckle nulling is a technique which modulates the electric field at a location in the focal plane with a DM to determine the settings required to cause destructive interference at that location. This reduces the star light and hence photon noise, reducing the required exposure time for detection. Speckle nulling can also be done through an optical fiber to minimize the amount of stellar leakage~\cite{Llop_DEF2019,LLop_LDW_2022}. With the availability of the new 1k DM, speckle nulling was implemented on KPIC. It utilized flux measurements made with NIRSPEC in quasi-real time and reduced the stellar leakage at 2~$\lambda/D$ by a factor of 2.6-2.8, limited by the variability of the wavefront, which changes significantly on the timescale of one control cycle~\cite{Xin_OSS_2023}. This has enabled KPIC to reach contrasts after post-processing of 1.3$\times$10$^{-4}$ at 90 mas separation from the star, assuming a planet atmosphere dominated by CO and water \cite{Wang2024_KPICContrastCurve}. This observing mode will be beneficial in the 2-5~$\lambda/D$ regime.

\subsection{Vortex Fiber Nulling} \label{sec:VFNDemo}
%Since the concept of vortex fiber nulling was established, it was clear that the relatively simple mode would be a good fit for KPIC~\cite{echeverri_VFN2019}, and could open a new parameter space of spectral characterization close to the optical axis. 
With the deployment of the charge 2 vortex mask in 2022, testing and commissioning swiftly followed. To learn more about the charge 2 VFN observing mode, on-sky performance and detection limitations please refer to~\cite{Echeverri_VFN2023}. The mode was used to detect several previously undetected close-in companions at moderate contrast~\cite{Echeverri_VFN2024} validating the approach and confirming the sensitivity limits.

In 2024, we deployed an additional charge 1 vortex mask as outlined in Sec.~\ref{sec:VFNDesign}. This mask was tested with on-sky observations against the charge 2 mask on the same night. The highlight of these tests was on 20$^{th}$ April 2024, when we observed HIP~62145. This is a previously-known binary star with an RV detection from SOPHIE~\cite{Kiefer2019_VFNHIP62145RV} and an astrometric detection from Gaia~\cite{Holl2022_GaiaNSS}, but no previous direct (i.e. imaging or interferometric) detection. We estimate the primary-to-secondary flux ratio at around 70 (contrast ${\sim}1.5{\times}10^{-2}$) based on the approximate companion mass from the published orbital solutions. There are large uncertainties in the orbital parameters from the previous detections, with the two solutions placing the secondary component around 35~mas to 60~mas from the primary, depending on which published parameters are used. This made it a great VFN commissioning target since VFN does not require precise knowledge of the companion position. For our commissioning tests, we observed with both the charge 1 and 2 vortex masks for about 35~min each to assess the relative performance of the two. 
%The companion had previously never been directly detected, and the orbit provided by the RV detections was poorly constrained such that the companion position and separation were poorly known on our observing night... Looking at the two likely solutions, the companion could have been at a separation of XX or YY~mas from the primary star. This made it a great VFN commissioning target. 

\begin{figure}[htbp]
    \centering\includegraphics[width = \linewidth]{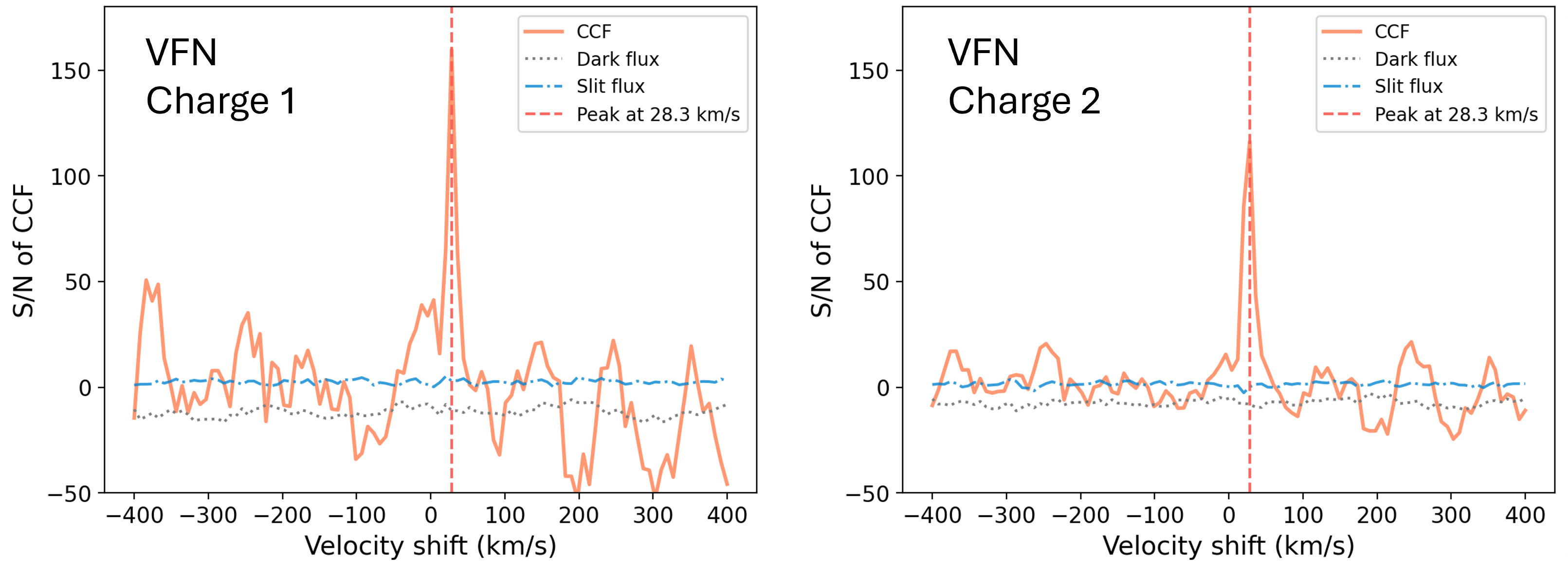}
    \caption{Demonstration of the new KPIC charge 1 VFN mode, alongside a re-validation of the extant charge 2 mode. Both plots show the CCF of a PHOENIX model template against 35~min of KPIC data on HIP~62145. The left plot is for the charge 1 VFN mode installed during the service mission, while the right plot is for the charge 2 VFN mode that had already been installed and demonstrated in 2022. \label{fig:VFNDetect}}
\end{figure}

We did a preliminary cross-correlation analysis on the resulting spectra using a PHOENIX model template~\cite{Husser2013_PHOENIX}, again assuming the expected companion mass from the published orbits. This yielded a strong detection in the KPIC data with both charge 1 and charge 2 VFN. Figure~\ref{fig:VFNDetect} shows the CCF between the model template and the KPIC data with both charge 1 (left) and charge 2 (right). The CCF peak for both VFN modes is at 28.3~km/s, which adds additional confidence in the reliability and consistency of the results between the two masks. Charge 1 has a stronger peak, indicating that the companion may have been at a smaller separation and also validating that despite the manufacturing challenges, the charge 1 mask is still beneficial for some targets. The detection strength in both CCFs is such that we are confident about the detection with both masks, but the results are still very preliminary since this is commissioning data. We haven't done any additional analyses or spectral retrievals to determine the companion composition, temperature, surface gravity, or any other properties. As demonstrated in a previous VFN paper~\cite{Echeverri_VFN2024}, all of these analyses are possible given the strength of the detection, but they would require additional analysis beyond the simple cross-correlation done here. However, the results here are only meant to show that the charge 1 mask has been installed, is working well, and provides a benefit over the charge 2 mask in specific cases.

\subsection{Laser Frequency Comb demonstration} \label{sec:LFCDemo}

\begin{figure}[htbp]
    \centering\includegraphics[width = 1.0\linewidth]{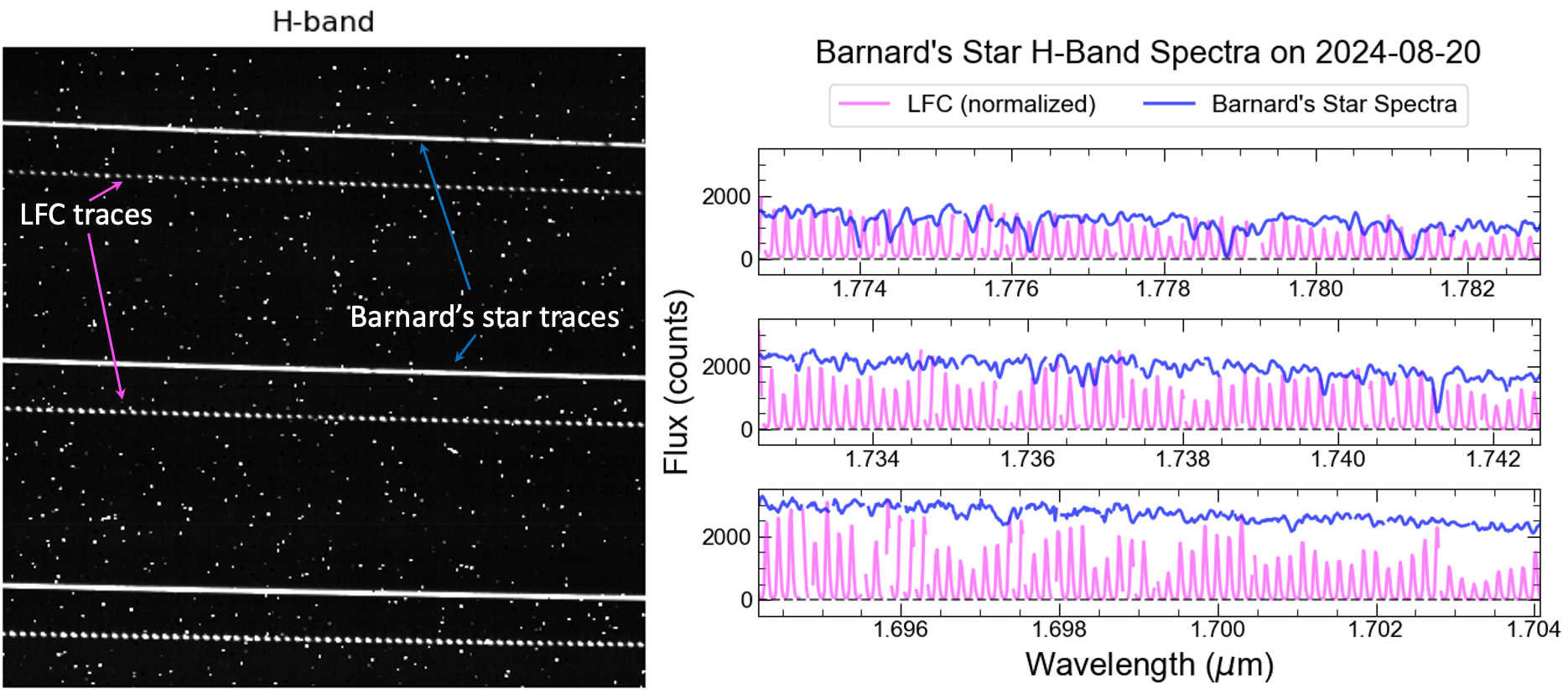}
    \caption{First on-sky KPIC spectra with the new HK LFC. (Left) This a small subwindow of the raw two-dimensional NIRSPEC frame at the long end of H band. There are two fiber traces per order. Blue arrows point to one of the science traces. Pink arrows point to one of the LFC traces. (Right) Extracted spectra showing the LFC trace in pink and the spectrum of Barnard's star in blue. In both the NIRSPEC frame and extracted spectra, the comb lines are clearly distinguishable. \label{fig:LFC}}
\end{figure}

Data were acquired with the LFC and Barnard's star on 2024-08-20 (UT). The observations were made with light from Barnard's star injected into the central science fiber for the yJH~bundle. The LFC was injected into one of the edge fibers on the bundle. Figure~\ref{fig:LFC} shows an example spectra acquired on-sky with the EO LFC and Barnard's star (uncorrected for telluric absorption) simultaneously in H~band. The LFC line intensity (counts), generally much stronger than the science spectra, are scaled to match the science spectra for better visualization and comparison by adjusting the VOA. In this case, nearly 50 dB of additional attenuation was needed for the LFC light in addition to the attenuation provided by the spectral flattener.
The left panel shows the raw two-dimensional spectral traces directly from the NIRSPEC detector. A small sub window with a portion of the echelle orders 43--45 is shown in the frame. The comb lines with their distinct periodic and well-defined structure are clearly visible in the LFC traces in each order. The right panel shows the extracted spectra for the LFC and Barnard's star for the region shown in the two-dimensional frame. Details about the precision of the LFC and stability of NIRSPEC are beyond the scope of this work, but  a preliminary analysis of the HK LFC data suggests on-sky wavelength stability $<$3 m s$^{-1}$ will be possible based on previous tests~\cite{Yi_DNI2016}, particularly in conjunction with KPIC's stable line spread function. This demonstration however validates simultaneous operations of the LFCs with science observations opening up this new observing mode. The primary limits to RV science with KPIC + LFC will be the deleterious effects of atmospheric absorption across the NIR bands and the limited cadence available within the overall Keck schedule.

\section{SUMMARY}
The Keck Planet Imager and Characterizer received a series of upgrades from 2022-2024 aimed at improving throughput, enhancing calibration efficacy and efficiency and reducing stellar leakage. The phase II version of the KPIC instrument has a throughput close to 5\% in good conditions which is almost twice that of the phase I variant. It can now support the spectroscopic characterization of companions within 1-2~$\lambda/D$ via the VFN mode in K band, provide improved contrast out to 5~$\lambda/D$ utilizing speckle nulling (a mode that is competitive with high contrast imagers like GPI \citenum{Wang2024_KPICContrastCurve}) and observe on or off-axis down to y band. All phase I capabilities have been recommissioned and the majority of phase II as well. The phase II development of KPIC has elucidated many practical solutions to for example static aberration compensation that can be leveraged by future single mode fiber instruments, such as HISPEC/MODHIS~\cite{Mawet_FFH2024}. KPIC is now in full scientific operation once again and is well positioned to build on the successes of the phase I variant. There will be no further upgrades to KPIC until it is decommissioned.

%%%%%%%%%%%%%%%%%%%%%%%%%%%%%%%%%
\section{Acknowledgements}
D. E. was supported by a NASA Future Investigators in NASA Earth and Space Science and Technology (FINESST) award \#80NSSC19K1423. 

K.H. is supported by the National Science Foundation Graduate Research Fellowship Program under Grant No. 2139433. 

J.X. is supported by the NASA Future Investigators in NASA Earth and Space Science and Technology (FINESST) award \#80NSSC23K1434.

Funding for KPIC has been provided by the California Institute of Technology, the Jet Propulsion Laboratory, the Heising-Simons Foundation (grants \#2015-129, \#2017-318, \#2019-1312, \#2023-4598), the Simons Foundation, and the NSF under grant AST-1611623.

The adaption of the Menlo comb for use with KPIC was funded by W. M. Keck Foundation under grant \#12540397. The construction of the HK EO comb was funded through a grant from the Heising-Simons Foundation grant \#2020-2399.

The W. M. Keck Observatory is operated as a scientific partnership among the California Institute of Technology, the University of California, and NASA. The Keck Observatory was made possible by the generous financial support of the W. M. Keck Foundation. 

Part of this work was conducted at the Jet Propulsion Laboratory, California Institute of Technology, under contract with NASA. 

We also wish to recognize the very important cultural role and reverence that the summit of Maunakea has always had within the indigenous Hawaiian community. We are most fortunate to have the opportunity to conduct observations from this mountain. 

%%%%% Data Availability Statement %%%%%
\section{Code and Data Availability Statement}
The data that support the findings of this article are not publicly available. They can be requested from the author at nem@caltech.edu.

%%%%% Disclosures %%%%%
\section{Disclosures}
The authors declare there are no financial interests, commercial affiliations, or other potential conflicts of interest that have influenced the objectivity of this research or the writing of this paper.

%%%%% References %%%%%

\bibliography{report}   % bibliography data in report.bib
\bibliographystyle{spiejour}   % makes bibtex use spiejour.bst

%%%%% Biographies of authors %%%%%

\vspace{1ex}
\noindent Biographies and photographs of the other authors are not available.

\listoffigures
\listoftables

\end{spacing}
\end{document}